\documentclass[amsmath,amssymb,showpacs,paper]{revtex4-1}
\usepackage{graphicx}

\def\rf#1{(\ref{#1})}
\def\lab#1{\label{#1}}
\def\nonu{\nonumber}
\def\br{\begin{eqnarray}}
\def\er{\end{eqnarray}}
\def\be{\begin{equation}}
\def\ee{\end{equation}}

\def\lb{\lbrack}
\def\rb{\rbrack}

\def\({\left(}
\def\){\right)}

\newcommand{\ct}[1]{\cite{#1}}
\newcommand{\bi}[1]{\bibitem{#1}}

\relax

\def\a{\alpha}

\def\b{\beta}

\def\d{\delta}

\def\g{\gamma}
\def\G{\Gamma}

\def\h{{1\over 2}}

\def\o{\over}

\def\pa{\partial}
\def\pr{\prime}

\def\ra{\rightarrow}
\def\s{\sigma}

\def\tp0{\Theta_{+}^{(0)}}
\def\tm0{\Theta_{-}^{(0)}}

\def\vp{\varphi}

\def\cl{{\cal L}}

\def\f#1#2#3 {f^{#1#2}_{#3}}

\def\win1{{\sf w_{1+\infty}}}

\def\Win1{{\sf W_{1+\infty}}}

\def\rlx{\relax\leavevmode}
\def\inbar{\vrule height1.5ex width.4pt depth0pt}
\def\IZ{\rlx\hbox{\sf Z\kern-.4em Z}}
\def\IR{\rlx\hbox{\rm I\kern-.18em R}}
\def\IC{\rlx\hbox{\,$\inbar\kern-.3em{\rm C}$}}
\def\IN{\rlx\hbox{\rm I\kern-.18em N}}
\def\IO{\rlx\hbox{\,$\inbar\kern-.3em{\rm O}$}}
\def\IP{\rlx\hbox{\rm I\kern-.18em P}}
\def\IQ{\rlx\hbox{\,$\inbar\kern-.3em{\rm Q}$}}
\def\IF{\rlx\hbox{\rm I\kern-.18em F}}
\def\IG{\rlx\hbox{\,$\inbar\kern-.3em{\rm G}$}}
\def\IH{\rlx\hbox{\rm I\kern-.18em H}}
\def\II{\rlx\hbox{\rm I\kern-.18em I}}
\def\IK{\rlx\hbox{\rm I\kern-.18em K}}
\def\IL{\rlx\hbox{\rm I\kern-.18em L}}
\def\one{\hbox{{1}\kern-.25em\hbox{l}}}
\def\0#1{\relax\ifmmode\mathaccent"7017{#1}%
B        \else\accent23#1\relax\fi}

\def\PRL#1#2#3{{\sl Phys. Rev. Lett.} {\bf#1} (#2) #3}
\def\NPB#1#2#3{{\sl Nucl. Phys.} {\bf B#1} (#2) #3}

\def\PRD#1#2#3{{\sl Phys. Rev.} {\bf D#1} (#2) #3}

\def\PLA#1#2#3{{\sl Phys. Lett.} {\bf #1A} (#2) #3}
\def\PLB#1#2#3{{\sl Phys. Lett.} {\bf #1B} (#2) #3}
\def\JMP#1#2#3{{\sl J. Math. Phys.} {\bf #1} (#2) #3}

\def\AoP#1#2#3{{\sl Ann. of Phys.} {\bf #1} (#2) #3}

\def\JHEP#1#2#3{{\sl JHEP} {\bf #1} (#2) #3}

\def\PAN#1#2#3{{\sl Physics of Atomic Nuclei} {\bf #1} (#2) #3}

\begin{document}
\begin{titlepage}
\vspace*{-1cm}

\noindent

\vskip 2cm

\vspace{.2in}
\begin{center}
{\large\bf A fermion-soliton system: self-consistent solutions, vacuum polarization and charge quantization}
\end{center}

\vspace{0.1in}

\begin{center}
Harold Blas

\vspace{.5 cm}
\small

\par \vskip .1in \noindent
Instituto de F\'{\i}sica\\
Universidade Federal de Mato Grosso\\
Av. Fernando Corr\^ea da Costa, N$^{0}_{-}$ 2367, Bairro Boa Esperan\c ca \\
78060-900, Cuiab\'a - MT - Brazil
\normalsize
\end{center}

\vspace{0.4in}

An integrable two-dimensional system related to certain fermion-soliton systems is studied. The self-consistent solutions of a static version of the system are obtained by using the tau function approach. The self-consistent solutions appear as an infinite number of topological sectors labeled by $n \in \mathbb{Z}_{+}$, such that in each sector the scalar field would evolve continuously from a trivial configuration to the one with half integer topological charge. The spinor bound states are found analytically for each topological configuration of the background scalar field. The bound state energy satisfies an algebraic equation of degree $2n$, so the study of the  energy spectrum finds a connection to the realm of algebraic geometry. We provide explicit computations for the topological sectors $n=1,2$. Then, by monitoring the energy spectrum, including the energy flow of any level across $E_n=0$, we discuss the vacuum polarization induced by the soliton. It is shown that the equivalence between the Noether and topological currents and the fact that the coupling constant is related to the one of the Wess-Zumino-Novikov-Witten (WZNW) model imply the quantization of the  spinor and topological charges. Moreover, we show that the soliton mass as a function of the boson mass agrees with the Skyrmes's phenomenological conjecture. Our analytical developments improve and generalize the recent numerical results in the literature performed for a closely related model by Shahkarami and  Gousheh, JHEP06(2011)116.  The construction of the bound states corresponding to the topological sectors $n \geq 3$ is briefly outlined.
 
\end{titlepage}

\section{Introduction}

Many aspects of soliton-fermion systems in two dimensions, such as the quantum corrections, have been studied in
 the literature; however, some important features at the zero order are yet to be investigated. In this context,
the $sl(2)$ conformal affine Toda model coupled to matter fields (CATM) is a two dimensional integrable and
conformally invariant field theory with spinor and scalar fields, which presents topological solitons in the
pseudoscalar field. It has two Dirac spinors coupled to a pseudoscalar field $\vp$ and a scalar field $\eta$ which
plays the role of a conformal symmetry `gauge' connection. By choosing a particular constant solution $\eta_o$
of the equations of motion the conformal symmetry is broken, thus
obtaining a massive theory, the so-called affine Toda model coupled
to matter (Dirac) fields (ATM) involving the Dirac spinors and the pseudoscalar field $\vp$ only.
The integrability of the theory is
established using a zero curvature formulation of its equations of motion
based on the ${\widehat {sl}}(2)$ affine Kac-Moody algebra. Some type of
solutions such as the one and two soliton solutions, associated to some particular vacua,
have been obtained in \ct{matter, npb1}. Besides the
conformal and local gauge symmetries, the model presents chiral symmetry and
some discrete symmetries. However, one of its main properties is that for certain solutions
there is an equivalence between the $U(1)$ Noether current, involving the spinors only, and the topological current
associated to the pseudoscalar soliton. That fact was established in
\ct{matter} at the classical level and in \cite{npb1} at the quantum mechanical level.
It implies that the density of the $U(1)$ charge has
to be concentrated in the regions where the pseudoscalar field has non vanishing
space derivative. The one-soliton of the
theory is a kink type soliton with fractional topological charge, and therefore the charge density is
concentrated inside the soliton. The CATM model constitutes an excellent laboratory
to test ideas about confinement \cite{npb1, npb2, prd1}, the role of solitons
in quantum field theories \cite{npb1} and duality transformations interchanging
solitons and particles \cite{npb2, aop1}.

A version of the off-critical ATM model with just one Dirac spinor has been considered in many physical situations related to  fermion-soliton systems. For example it
appears in the study of the dynamical Peierls' energy gap generation in one-dimensional
charge-density wave systems \cite{belve1}, quantum field-theory description of tunneling in
the integer quantum Hall effect \cite{mori1}, discussions of fractional charge or topological charges  induced
in the ground state of a Fermi system by its coupling to a chiral field
\cite{goldstone, keil,stone,gousheh, mackenzie, polychronakos}, and  the spectrum
and string tension of the low-energy effective Lagrangian of QCD$_2$ \cite{prd1}. The model has also been considered  in order to discuss the realization of chiral symmetries in $(1+1)$ dimensions \cite{witten}. In the strong coupling limit it is closely related to the integrable variable mass sine-Gordon model presented in \cite{kundu, uni1}. Recently, a self-consistent
solutions have been considered for a related version of the model with additional $\vp^4$ potential
using a numerical method \cite{Shahkarami1} and the Casimir energy of that system has been computed, such that a static  pseudoscalar
field is prescribed with two adjustable parameters \cite{Shahkarami2}.

Recently, there has been significant growth in the study of  fermion-soliton systems in the context of
the localization and bound states
of the fermion on the
so-called braneworld models, following the original work of Rubakov and Shaposhnikov \cite{rubakov}. There are some
models in which the brane Lagrangian considers the kink soliton
of the
$\vp^4$ or sine-Gordon type systems in one dimension (see e.g. \cite{liu}). Some related issues such as the
particle transfer and fermion spectrum in braneworld collisions have also been considered \cite{saffin, gibbons}.

In \cite{npb1} the soliton type solutions of the ATM model associated to some type of vacua have been obtained
using a hybrid of the dressing transformation and the Hirota's
method. The basic idea was to look for vacuum
configurations where the Lax potentials lie in an abelian (up to central terms)
subalgebra of $\widehat{sl} (2)$. The solitons are obtained by performing the
dressing tranformations from those vacuum configurations. Such procedure led  quite
naturally to the definition of tau-functions, and then the Hirota's
method was easily implemented too. The authors  discussed the conditions for the
solutions to be real, and evaluated the topological charges. The interactions of
the solitons was studied by calculating their time delays. It has been shown that it is
attractive, and the time delays have shown to be the same as those for the
sine-Gordon solitons. An interesting aspect of the  solutions is that when the
two Dirac spinors of the theory are  related by a reality
condition, then either the soliton or the anti-soliton disappears from the
spectrum. It has been  interpreted as  indicative of the existence of a duality
involving the solitons and the spinor particles. In addition, the system can be mapped to either,
the sine-Gordon model or the massive Thirring model through different gauge fixing procedures. In that scenario the solitons of the sine-Gordon model are interpreted as the spinor particles of the
Thirring theory. Thus, the ATM theory is related to the sine-Gordon and the massive Thirring models in its strong/weak coupling sectors, respectively \cite{npb2, aop1}.

An important property of
the model under consideration is that it possesses special type of scalar and spinor fields in the intermediate field configurations in which they interact. In this scenario the role played by the local gauge symmetries on the physical soliton spectrum, as well as the Noether and topological currents equivalence deserve some clarifications. Here we show that the scalar field  can represent topological solitons with infinitely many different shapes  and the corresponding fermion becomes bound state solutions. The  topological charge can take any real value  as a consequence of the fact that
the ATM model does not possess a particle conjugation symmetry \cite{goldstone}.  In the intermediate region of field space in which the spinor and scalar fields interact to greater or
lesser extent the system possesses rich physical properties. The choice of complicated background configurations 
associated to the scalar field renders the problem analytically unsolvable. Then one could 
resort to certain numerical solutions, in which case some important physical properties and concepts  might not be clear. Some approaches have been proposed to circumvent this difficulty, they regard the scalar field as an external potential with prescribed shape, e.g. in \cite{mackenzie, polychronakos} it is assumed as an external infinitely thin (wide) soliton, whereas in  \cite{keil} it is assumed to be a  piecewise linear potential, allowing a set of analytic solutions for the spinor bound states \cite{gousheh}. 

In this paper we provide self-consistent solutions for a version of the ATM model with one spinor field coupled to the pseudoscalar, as considered in the literature. The static version of the system of equations allows a set of fermionic bound states corresponding to a family of soliton profiles for the scalar field. The self-consistent
solutions appear as an infinite number of topological sectors labeled by an positive integer $n \in \mathbb{Z}_{+}$, such that in each sector the scalar field would evolve continuously from a trivial configuration to the one with topological charge $Q_{top} = \pm \frac{n}{2}$. The bound states are found analytically for each topological configuration of the background field. The potential functional related to the scalar field is
generated dynamically and its expansion would
reproduce the scalar particle mass and $\Phi^4$ theory interaction terms. We obtain the energy levels of the bound states and investigate the relationship between the soliton, fermion and boson masses in each sector. These relationships will be used to discuss the Skyrme conjecture regarding the
ratio between the boson and soliton masses. Recently, 
in \cite{Shahkarami1} the authors studied, by means of a numerical method, an analogous model by incorporating a scalar particle mass term and the $\Phi^4$ theory potential by hand. Our analytical calculations are compared to some of the numerical results of that reference, providing exact corrections to some quantities and identifying certain qualitative resemblances.  

We perform explicit computations of the bound state energies $E_{n}$ for the topological sectors labeled by $n=1,2\,$ as the background field evolves from a topological trivial configuration to a nontrivial one, for various soliton and spinor bound state parameters. The steps followed in order to construct the bound state energy $E_n$ corresponding to the sectors $n \geq 3$ will be  outlined.  Then, by monitoring the energy spectrum, including the energy flow of any level across $E_n=0\, (n=1,2 )$, we discuss the vacuum polarization induced by the soliton. The spinor charge quantization is discussed in view of the equivalence between the Noether and topological currents and the fact that the ATM coupling constant is related to the integer coupling constant of the WZNW model.  

The paper is organized as follows: section \ref{sec:model} summarizes the main properties of
the ATM model. In section \ref{sec:topo} a new set of fields are defined: a scalar field $\Phi$ associated to topological configurations and some  spinor components $\xi_{a}\, (a=1,2,3,4)$ related to the fermion bound states. In section \ref{sec:tau} the self-consistent solutions are obtained through the tau function approach. In subsections \ref{subsec:n1} and \ref{subsec:n2} we perform explicit calculations corresponding to the topological sectors $n=1$ and $n=2$, respectively. The solitons and the corresponding spinor bound states are computed, as well as the bound state energies and the soliton masses in each case. The boson mass associated to the scalar $\Phi_n \, (n=1,2)$ is computed providing the same formula for the both sectors $n=1,2$.  In the both cases the bound state energy $E_n$ is a solution of a relevant  algebraic equation. In section \ref{sec:pol} the vacuum polarization of the fermion induced by the soliton is discussed for any sector $n\in \mathbb{Z}_{+}$.  In section
\ref{sec:duality} we discuss some properties of the model, such as the currents equivalence and charge quantization. In   section  \ref{sec:ratio} the  ratio (soliton mass)/(boson mass)  is discussed for the sectors $n=1,2$. In the  section 
  \ref{sec:n3} it  is provided a brief discussion on the bound states related to the topological sectors $n\geq 3$ and in section \ref{sec:discussion} it is presented  the discussions of the main results. The paper  presents 20 figures in order to illustrate the solutions and main results. The appendix 
\ref{app:norman2} presents some  results concerning the topological sector $n=2$.

\section{The model}
\label{sec:model}

Consider the two-dimensional field theory defined by the Lagrangian 
\br
\cl =  -{1\o 2} \pa_{\mu} \vp \, \pa^{\mu} \vp
 + i  {\bar{\psi}} \gamma^{\mu} \pa_{\mu} \psi
- m_{\psi}\,  {\bar{\psi}} \,
e^{\eta+i\sqrt{8/k}\, \vp\,\gamma_5}\, \psi + \h  \pa_{\mu} \nu \, \pa^{\mu} \eta
- {k\o 8}\, m_{\psi}^2 \, e^{2\,\eta}
\label{lagrangian1}
\er
where $\vp$, $\eta$ and $\nu$ are scalar fields, and $\psi$ is a
Dirac spinor. It is defined ${\bar{\psi}} \equiv {\widetilde \psi}^{T} \,
\gamma_0$, with ${\widetilde \psi}$ being a second Dirac spinor. The $sl(2)$ CATM Lagrangian (\ref{lagrangian1}) follows
 if one makes  $\vp \rightarrow \sqrt{\frac{2}{k}}\,\, \vp,\,\nu \rightarrow \frac{1}{k}\,\, \nu,\,\psi \rightarrow
\sqrt{\frac{1}{k}}\,\, \psi,\,\widetilde{\psi} \rightarrow \sqrt{\frac{1}{k}}\,\, \widetilde{\psi}$
in the model of \cite{npb1}. The CATM model is related to the so-called two-loop Wess-Zumino-Novikov-Witten (WZNW) model through a Hamiltonian reduction such that  $k=\frac{q}{2\pi},\,q \in \mathbb{Z}$, where $q\in \mathbb{Z}$ is the WZNW coupling constant. In many applications of \cite{npb1} it has been considered the reality condition
\be
{\widetilde \psi} = e_{\psi} \,\psi^*
\label{psitrealcond}
\ee
where $e_{\psi}$ is a real constant. It has been shown that the sign of
$e_{\psi}$ plays  an important role in  determining the spectrum of
soliton solutions. The theory (\ref{lagrangian1}) in \cite{matter} is
defined for a general complex fields, i.e. a complex Lagrangian. In order to define a suitable physical Lagrangian one can follow  the prescription to restrict the model
to a subspace of well-behaved classical solutions \cite{aop1}. For example, for $e_{\psi}=-1$ one can drop an overall minus sign from the Lagrangian in order to get usual kinetic terms for the spinor and $\vp$ scalar fields in  such a way that one defines  a submodel with a Hamiltonian
bounded from below. In fact, as presented in \cite{npb1} the one and two
soliton solutions for $\vp$ and the corresponding spinor solutions satisfy the reality condition (\ref{psitrealcond}).

One of the main conclusions in ref. \cite{npb1} is that without the reality condition (\ref{psitrealcond}) the theory
(\ref{lagrangian1}) has two Dirac spinors and it also has the soliton and anti-soliton solutions associated to $\vp$, since both signs of the soliton charge are admissible. However, if one  imposes the
reality condition (\ref{psitrealcond})  the theory (\ref{lagrangian1}) looses one Dirac spinor and also one-soliton (two-soliton) solution, since for a given choice of $e_{\psi}$ only one topological charge $Q_{topol}= \mbox{sign}\,\(e_{\psi}\)\,\,\frac{1}{2}$ [$Q_{topol}= \,\,\mbox{sign}\,\(e_{\psi}\)$ ]
corresponding to one-soliton (two-soliton) or one-anti-soliton (two-anti-soliton), is permitted. That is indicative of the existence
of some sort of duality between solitons and antisolitons associated to $\vp$ and the spinor particles $\psi,\,\, \widetilde{\psi}$, respectively. So, one can have (without (\ref{psitrealcond})) soliton and anti-soliton solutions with $\vp$ real. However, a charge zero solution corresponding to the scattering of a soliton
and anti-soliton for $\vp$ real does not exist. Such solution exists, however, for $\vp$ complex but asymptotically
real.

The corresponding equations of motion are  
\br
\pa^2 \,\vp &=& i m_{\psi}\, \sqrt{\frac{8}{k}}\,\, \overline
\psi \gamma_5 e^{\eta+i \sqrt{\frac{8}{k}}\,\varphi \gamma_5} \psi,
\lab{sl2eqm1} \\
i \gamma^{\mu} \pa_{\mu} \psi &=& m_{\psi}\, e^{\eta+i\sqrt{\frac{8}{k}} \,\vp\,\gamma_5} \,
\psi ,\lab{sl2eqm4}\\
i \gamma^{\mu} \pa_{\mu} \widetilde \psi &
=& m_{\psi}\, e^{\eta - i \sqrt{\frac{8}{k}} \vp\,\gamma_5} \,
\widetilde  \psi ,
 \lab{sl2eqm5}\\
 \pa^2\,{\nu} &=& - 2 m_{\psi}\,  \overline
\psi e^{\eta+i \sqrt{\frac{8}{k}}\, \varphi \gamma_5} \psi
 - {k \over 2}  m_{\psi}^2 e^{2\eta},
\lab{sl2eqm2}\\
\pa^2 \eta &=& 0,
\lab{sl2eqm3}.
\er

The integrability properties, the construction of soliton 
solutions for $\eta, \vp , \psi$ fields associated to a particular vacua, and many other properties were discussed
in \ct{matter, npb1}. We will be interested in special type of topological configurations associated to a scalar field $\Phi$ and their relevant fermionic bound states $\chi_a$. These new fields will be related to the original $\vp$ and $\psi, \widetilde{\psi}$ fields, respectively, by means of certain transformations. We start by reviewing the
symmetries of \rf{lagrangian1} and the eqs. of motion (\ref{sl2eqm1})-(\ref{sl2eqm3}).

{\em Left-right local symmetries}. The Lagrangian \rf{lagrangian1} is invariant 
under the local $U(1)_L \otimes U(1)_R$ transformations
\be
\vp \ra \vp + \sqrt{\frac{k}{2}}\,\varepsilon_{R}\( x_{+}\) + \sqrt{\frac{k}{2}} \, \varepsilon_{L}\( x_{-}\) \; ; \qquad
\nu \ra \nu \; ; \qquad \eta \ra \eta \label{lr1}
\ee
and
\be
\psi \ra e^{- i\,[\( 1+ \gamma_5\) \varepsilon_{R}\( x_{+}\)
- \( 1- \gamma_5\)  \varepsilon_{L}\( x_{-}\)]}\, \psi
\; ; \qquad
{\widetilde \psi} \ra e^{ i\,[\( 1+ \gamma_5\) \varepsilon_{R}\( x_{+}\)
- \( 1- \gamma_5\) \varepsilon_{L}\( x_{-}\)]} {\widetilde \psi} \label{lr2}
\ee

Notice that the eqs. of motion are invariant under these transformations, whereas the Lagrangian (\ref{lagrangian1}) is invariant up to a surface term.

{\em $U(1)$ global symmetry}. Notice that, by taking $\varepsilon_{R}\( x_{+}\) = -
\varepsilon_{L}\( x_{-}\) = - \h \,\,
\theta$, with $\theta = {\rm const.}$, one gets a global $U(1)$ transformation
\be
\vp \ra \vp  \; ; \qquad
\nu \ra \nu \; ; \qquad \eta \ra \eta   \; ; \qquad
\psi \ra e^{i \theta} \, \psi  \; ; \qquad
{\widetilde \psi} \ra e^{-i \theta} \, {\widetilde \psi}
\lab{globalu1}
\ee
The corresponding Noether current is given by
\be
J^{\mu} = \frac{1}{k} {\bar{\psi}}\, \gamma^{\mu}\, \psi \, , \qquad
\pa_{\mu}\, J^{\mu} = 0.
\lab{noethersl2}
\ee
Let us define the $U(1)$ charge as
\br
Q_{\psi}  = \frac{1}{k \pi} \int_{-\infty}^{+ \infty}  dx \,  J^{0},\label{psicharge}
\er 
where the normalization factor $1/\pi$ has been introduced for later convenience.

{\em Chiral symmetry}. In addition, if one takes $\varepsilon_{R}\( x_{+}\) =
\varepsilon_{L}\( x_{-}\) = - \h \,\,  \a$,  with $\a = {\rm const.}$, one gets  the global
chiral symmetry
\be
\psi \ra e^{i\gamma_5 \a}\, \psi \; ; \qquad
{\widetilde \psi}\ra e^{-i\gamma_5 \a}\, {\widetilde \psi} \; ; \qquad
\vp \ra \vp -  \a \; ; \qquad
\nu \ra \nu \; ; \qquad \eta \ra \eta
\lab{chiraltransf}
\ee
with the corresponding Noether current
\be
J_5^{\mu} =  \frac{1}{k} \bar \psi \gamma_5 \gamma^\mu \psi
+{1\over 2} \sqrt{\frac{2}{k}}\, \partial^\mu \vp
\; ; \qquad \qquad \pa_{\mu}J_5^{\mu} =0
\lab{chiral}
\ee

{\em Topological charge}. One can shift the $\vp$ field as $\vp \ra \vp + n_1
\pi \,\sqrt{\frac{k}{2}}\,\, (n_1 \in \mathbb{Z})$, keeping all the other fields unchanged, that the Lagrangian is left
invariant. That means that the  theory possesses an infinite number of vacua,
and the topological charge
\be
Q_{\rm topol.} \equiv \int \, dx \, j^0
\, , \qquad
j^{\mu} =  {1\o{ 2 \pi }} \sqrt{\frac{2}{k}}\,\epsilon^{\mu\nu} \pa_{\nu} \, \vp
\lab{topological}
\ee
can assume non trivial values. This model possesses fractional topological charges as we will see below.

{\em  CP-like symmetry}. In addition,  the Lagrangian \rf{lagrangian1} is invariant
under the transformation
\be
x_{+} \leftrightarrow x_{-} \, ; \quad \psi
\leftrightarrow i\epsilon  \gamma_0{\widetilde {\psi}} \, ; \quad
{\widetilde {\psi}} \leftrightarrow - i\epsilon   \gamma_0 \psi \, ; \quad
\vp \leftrightarrow \vp \, ; \quad
\eta \leftrightarrow \eta \, ; \quad
\nu \leftrightarrow \nu
\lab{cpsymmetry}
\ee
where $\epsilon = \pm 1$. The reality condition
\rf{psitrealcond} breaks such CP symmetry, for any real value of the
constant $e_{\psi}$. The CP symmetry is preserved if one
takes $e_{\psi}=\pm i$. In addition, the theory (\ref{lagrangian1}) is not invariant under spatial parity. In general, solitons are transformed into anti-solitons by parity transformations; so, it is very unlikely that the theory has the both type of solitons associated to $\vp$ for a given choice of $e_{\psi}$ in (\ref{psitrealcond}). However, maintaining the constraint (\ref{psitrealcond}) the both type of solitons arise for certain submodels defined for some gauge invariant fields, which turn out to be the sine-Gordon and the massive Thirring models, in the scalar and the spinor sectors of the ATM model, respectively \cite{aop1, npb2, jhep1, jhep11}.

{\em Conformal symmetry}. The model  \rf{lagrangian1} is invariant under the
conformal transformations (We are using $x_{\pm}=t \pm x$, and so,
$\pa_{\pm} = \h \( \pa_t \pm \pa_x\)$, and
$\pa^2 = \pa_t^2 - \pa_x^2 = 4 \pa_{+}\pa_{-}$.)
\be
x_{+} \ra {\hat x}_{+} = f(x_{+}) \, , \qquad
x_{-} \ra {\hat x}_{-} = g(x_{-}),
\lab{ct}
\ee
with $f$ and $g$ being analytic functions; and with the fields transforming as
\br
\vp (x_{+}\, , \, x_{-}) &\ra&
{\hat {\vp}}({\hat x}_{+}\, , \,  {\hat x}_{-}) =
\vp (x_{+}\, , \, x_{-}) \, ,
\nonu\\
e^{-\nu (x_{+}\, , \, x_{-})} &\ra&
e^{-{\hat \nu}({\hat x}_{+}\, , \,
{\hat x}_{-})} = \( f^{\pr}\)^{\d} \, \( g^{\pr}\)^{\d}
e^{-\nu (x_{+}\, , \, x_{-})} \, ,
\lab{ctf}\\
e^{-\eta (x_{+}\, , \, x_{-})} &\ra& e^{-{\hat \eta}({\hat x}_{+}\, , \,
{\hat x}_{-})} = \( f^{\pr}\)^{\h} \, \( g^{\pr}\)^{\h}  e^{-\eta (x_{+}\,
, \, x_{-})} \, ,
\nonu\\
\psi (x_{+}\, , \, x_{-}) &\ra & {\hat {\psi}} ({\hat x}_{+}\, , \,
{\hat x}_{-}) =   e^{{1\o 2}\( 1+ \gamma_5\) \log \( f^{\pr}\)^{-\h}
+ {1\o 2}\( 1- \gamma_5\) \log \( g^{\pr}\)^{-\h}}
\, \psi (x_{+}\, , \, x_{-}) \, ,
\nonu
\er
where the conformal weight $\d$, associated to $e^{-\nu}$, is arbitrary, and
$\widetilde \psi$ transforms in the same way as $\psi$.

Associated to the conformal symmetry \rf{ct} there are two chiral currents
\be
{\cal J}= - i\frac{1}{k} {\widetilde {\psi}}^T \( 1+\gamma_5 \)\psi
+ \sqrt{\frac{2}{k}}\, \partial_+\vp +\partial_+\eta, \qquad
{\bar {\cal J}}=  i \frac{1}{k} {\widetilde {\psi}}^T \( 1-\gamma_5 \)\psi
+ \sqrt{\frac{2}{k}}\, \partial_-\vp+ \partial_-\eta
\lab{chiralcur}
\ee
satisfying
\be
\partial_-{\cal J}=0 \; ; \qquad \quad  \partial_+{\bar {\cal J}}=0.
\ee

Observe, from \rf{ctf}, that the currents ${\cal J}$ and ${\bar {\cal J}}$ have
conformal weights $(1,0)$ and $(0,1)$ respectively. Under the conformal transformations \rf{ct}, the chiral currents transform as
\br
\lab{chi1}
\;\;{\cal J}(x_{+})&\longrightarrow& [\ln f^{\pr}( x_{+})
]^{-1}\left({\cal J}(x_{+})-[\ln f^{\pr}(x_{+})]^{\pr}\right),\\
\lab{chi2}
\overline{{\cal J}}(x_{-})&\longrightarrow& [\ln g^{\pr}(x_{-})
]^{-1}\left(\overline{{\cal J}}(x_{-})-[\ln
g^{\pr}( x_{-})]^{\pr}\right)
\er

Then, given a solution of the model, one can always map it, under a conformal transformation, into a solution where
\br
{\cal J}=0 \; ; \qquad \quad {\bar {\cal J}}=0.
\lab{constraints}
\er

Such a procedure amounts to gauging away the free field $\eta$ \cite{aop1}. In other words, (\ref{constraints}) are constraints implementing a Hamiltonian reduction. So, for every regular solution $\eta \equiv \bar{\eta}$ the CATM defined on a space-time $(x_+,x_-)$ will correspond  to an off-critical submodel, the so-called affine Toda model coupled to matter (ATM). For  the particular solution $\bar{\eta} = 0$ the CATM and the ATM are defined on the same space-time. So, for $\eta=\eta_o=constant$\, the Lagrangian for the set of equations of motion corresponding to the Toda and Dirac fields becomes
\br
\cl =  -{1\o 2} \pa_{\mu} \vp \, \pa^{\mu} \vp
 + i  {\bar{\psi}} \gamma^{\mu} \pa_{\mu} \psi
- m_{\psi}\, e^{\eta_o} \,{\bar{\psi}} \,
e^{i\sqrt{8/k}\, \vp\,\gamma_5}\, \psi.
\lab{atm1}
\er

This Lagrangian is not conformal invariant and defines the off-critical ATM model which we will consider in this work. One can easily check that the constraints \rf{constraints}, once we have set $\eta=\eta_0$, are equivalent to
\be
{1\o{2 \pi }} \sqrt{\frac{2}{k}}\,\, \epsilon^{\mu\nu} \pa_{\nu} \, \vp =
{1\o k \pi} \bar \psi \gamma^\mu  \psi.
\lab{equivcurrents}
\ee

Therefore, in the reduced model, the Noether  current \rf{noethersl2} is
proportional to the topological current \rf{topological}. That fact has profound consequences in the properties of such theory. For instance, it implies (taking ${\widetilde {\psi}}$ to be proportional to the complex
conjugate of $\psi$) that the charge density $\psi^{\dagger}\psi$ is
proportional to the space derivative of $\vp$. Consequently, the Dirac field
is confined to live in regions where the field $\vp$ is not constant. The best
example of that is the
one-soliton solution of \rf{lagrangian1} which was calculated in \ct{matter}
and it is given by
\br
\vp &=& 2 \sqrt{\frac{k}{2}} \arctan \( \exp \( 2m_\psi \( x-x_0-vt\)/\sqrt{1-v^2}\)\) \label{sol1} \\
\psi &=& \sqrt{k} e^{i\theta} \sqrt{m_\psi} \,
e^{m_\psi \( x-x_0-vt\)/\sqrt{1-v^2}}\, \(
\begin{array}{c}
\left( { 1-v\o 1+v}\right)^{1/4}
{1 \o 1 + i e^{2 m_\psi \( x-x_0-vt\)/\sqrt{1-v^2}}}\\
-\left( { 1+v\o 1-v}\right)^{1/4}
{1 \o 1 - i e^{2 m_\psi \( x-x_0-vt\)/\sqrt{1-v^2}}}
\end{array}\)
\label{sol11}\\
\nu &=&
- \frac{k}{2} \log \( 1 + \exp \( 4 m_\psi \( x-x_0-vt\)/\sqrt{1-v^2}\)\)
- {k\o 8} m_{\psi}^2 x_{+} x_{-}
\label{sol111}\\
\eta & = & 0
\label{solsimple}
\er
and the solution for $\widetilde \psi$ is the complex conjugate of
$\psi$, implying $e_{\psi}=1$ in (\ref{psitrealcond}). Notice that, from \rf{topological}, one has $Q_{\rm topol.} =\frac{1}{2}$
for the  solution \rf{sol1}. In fact, this is a fractional charge, which is in accordance to the topological charge definition provided in \cite{matter}.

Notice that the solution for $\vp$ is of the kink type soliton, and
therefore $\pa_x \vp$ is non-vanishing only in a region of size of the order
of $m_\psi^{-1}$. In addition, the solution for $\psi$ resembles the massive
Thirring model type soliton.
One can check that (\ref{sol1})-(\ref{sol111}) satisfy
\rf{equivcurrents}, and so is a solution of the reduced model. However, one must emphasize that these solutions are not truly sine-Gordon/massive Thirring solitons, e.g. its topological charge is $+1/2$ whereas the $1-$soliton of the usual sine-Gordon model possesses integer charge $+1$.

We point out that the condition \rf{equivcurrents} together with the equations
of motion for the Dirac spinors \rf{sl2eqm4}-\rf{sl2eqm5} imply the equation
of motion for $\vp$, namely \rf{sl2eqm1}. Therefore in the reduced model,
defined by the constraints \rf{constraints}, one can replace a second order
differential equation, i.e. \rf{sl2eqm1}, by two first order equations,
i.e. \rf{equivcurrents}.

The model (\ref{atm1}) with additional mass and $\vp^4$ interaction terms has recently been
 considered in \cite{Shahkarami1, Shahkarami2}, and it has been solved by means of a numerical method and the
bound states of the fermion and the shape of the
soliton have been obtained. In this type of model the pseudoscalar field boundary condition is {\sl a priori}
prescribed since the zeros of the potential provides the values the $\vp$ field gets at spatial infinity, in
order to achieve finite energy and localized energy density solitons. Whereas, in the ATM model the scalar field
soliton acquires, apart from $n_{1} \pi$, new boundary conditions compatible with the dynamics, as will be seen below. The topologically nontrivial configuration of the scalar field $\vp$ of the ATM model, which
 is coupled to the ATM spinor field, can be thought to evolve continuously and slowly from a topologically trivial
 configuration to get a soliton configuration by a continuous gauge transformation, and the corresponding spinor bound states  get a continuous set of configurations. Then the topological charge 
of the intermediate solitons associated to the field $\vp$ can get any real value. This is reminiscent of the ideas present in the so-called smooth bosonization method \cite{damgaard} in which the bosonic and fermionic formulations of the same theory is understood as two different gauge fixings of a larger gauge-invariant action containing the both fermionic and bosonic fields. Equivalence of the both sectors of the theory then amount to the usual gauge-fixing independence of the S-matrix elements in gauge theories. This method has been used to derive the bosonization rules for certain models. It implies that the standard bosonization and fermionization
prescriptions interpolate two extreme field representations of a ``larger" theory which
in a completely smooth way can be brought to almost any desired form in between. The
equivalence between certain bosonic and fermionic theories (e.g. sine-Gordon/massive Thirring) then has profound consequences in the properties of the original gauge invariant theory, extending
to a continuum of theories in which fermions and bosons appear interactively to a greater
or lesser extent.  

\section{Topological configurations coupled to the spinors $\xi_{j}$}
\label{sec:topo}

In the next steps we rewrite the eqs. of motion of the ATM model in terms of new real scalar fields $\Phi,\,\zeta_1, \zeta_2$ and real spinor fields $\xi_{a}\,\,(a=1,...,4)$  defined by the relationships 
\br
\psi_L & = & e^{-i\zeta_{2}/2}\(\xi_1 + i\xi_2\),\,\,\,
\psi_R  = e^{i \zeta_{1}/2} \(\xi_3 +i \xi_4\),\label{zetas}\\
\Phi &=&  \sqrt{8/k}\, \vp +(\zeta_1+\zeta_2)/2\label{zetas1}
\er
 
Our aim is to investigate certain intermediate configurations coupling the scalar $\Phi$ and the fermion field, such that any desired nontrivial topological configurations can be achieved by the field $\Phi$. It is expected that the spinors  $\xi_a$ and the scalar $\Phi$ will be the analogs of the massive Thirring and the sine-Gordon fields, respectively. So, the system (\ref{sl2eqm1})-(\ref{sl2eqm5}), provided that $\eta=\eta_0=const.$, becomes
\br
\label{521c}
2\partial_+\xi_1+\partial_+ \zeta_2\,\, \xi_2 - M_{1} \xi_4 \sin{\Phi}+ M_{1} \xi_3 \cos{\Phi}&=&0,    \\
 \label{522}
2\partial_+\xi_2-  \partial_+\zeta_2 \,\, \xi_1+M_{1} \xi_3 \sin{\Phi}+ M_{1} \xi_4\cos{\Phi}&=&0, \\ \label{523}
2\partial_-\xi_3- \partial_-\zeta_1\,\, \xi_4 -M_{1} \xi_2 \sin{\Phi}- M_{1} \xi_1\cos{\Phi}&=&0, \\ \label{524}
2\partial_-\xi_4+ \partial_-\zeta_1\,\, \xi_3+M_{1} \xi_1\sin{\Phi}- M_{1} \xi_2\cos{\Phi}&=&0,  \\
\frac{16 M_{1}}{k} \Big[ (\xi_1\xi_3+\xi_2\xi_4)\cos{\Phi}- (\xi_1\xi_4-\xi_2\xi_3)\sin{\Phi}\Big]-&&
\nonumber\\
\partial^2\Phi+\frac{1}{2} \partial^2 \zeta_1 +\frac{1}{2} \partial^2 \zeta_2] &=&U_1'[\Phi],\,\,\,\,\,M_{1} \equiv e_{\psi} m_{\psi}\, e^{\eta_o}. \label{525} 
\er

The appearance of the term involving $U_1[\Phi]$ in the eq. (\ref{525}) deserves some clarifications. First,
since the  model possesses a local gauge invariance one needs to introduce a certain gauge fixing condition, so
 this term  would play this role. Second, this term will be part of the potential for the scalar $\Phi$,
 providing the vacuum points in order to find localized soliton solutions as in \cite{Shahkarami1}. Third,
in the static version of the eq. (\ref{525}), the interplay between the functional $U_1[\Phi]$ and
the combination $(\zeta_1+\zeta_2)$ will reproduce an equation invariant under parity, as will be seen below.

The static version of the system (\ref{521c})-(\ref{525}) written in the form of an eigenfunction equation for the Hamiltonian, with relevant mass and interaction terms for a $\Phi^4$ model, has been solved in \cite{Shahkarami1} by numerical methods. In fact, setting $\zeta_1 \equiv E\, x_{-} + h_{+}(x_{+}),\,\, \zeta_2 \equiv -E \,x_{+} + h_{-}(x_{-})$ and taking  the static version of the system  (\ref{521c})-(\ref{525}) one gets   
\br\label{5211}
\xi^{'}_1 - E\, \xi_2 - M_{1} \xi_4 \sin{\Phi}+ M_{1} \xi_3 \cos{\Phi}&=&0, \\ \label{5221}
\xi^{'}_2 + E\, \xi_1+ M_{1} \xi_3 \sin{\Phi}+ M_{1} \xi_4\cos{\Phi}&=&0, \\ \label{5231}
\xi^{'}_3+ E\, \xi_4 + M_{1} \xi_2 \sin{\Phi}+ M_{1} \xi_1\cos{\Phi}&=&0, \\ \label{5241}
\xi^{'}_4 -E\,  \xi_3-M_{1} \xi_1\sin{\Phi}+ M_{1} \xi_2\cos{\Phi}&=&0, \\
\frac{16 M_{1} }{k} \Big[ (\xi_1\xi_3+\xi_2\xi_4)\cos{\Phi}- (\xi_1\xi_4-\xi_2\xi_3)\sin{\Phi}\Big] + \Phi^{''} -(\bar{\zeta})^{''} &=&U_1'[\Phi],\label{phizetas1}
\er
where $E$ is the eigenvalue in the spinor sector and $\bar{\zeta} \equiv(\zeta_1+\zeta_2)/2$. The role played by the auxiliary field $\bar{\zeta}$ will be
discussed below in more detail. In the static version of the ATM system (\ref{5211})-(\ref{phizetas1}) the functional
 $U_1[\Phi]$ must be chosen such that, for every set of solutions
$\{ \Phi,\, \xi_a, \, (a=1,2,..4) \} $ satisfying the system (\ref{5211})-(\ref{5241}),
the kink would remain as a self-consistent solution of the eq. (\ref{phizetas1}). It is useful to consider the
spinor components  $\xi_a[\Phi]$ and the auxiliary field $\bar{\zeta}[\Phi]$ to depend functionally on $\Phi$ as discussed in \cite{dashen} regarding  the  self-consistent solutions of the $\Phi^4$ theory coupled to a fermion. Recently, the problem of a self-consistent solution for a
fermion coupled to static scalar field in the form of a kink (domain wall) \cite{Shahkarami1, gani} has been considered
mainly using numerical methods.
The self-consistent approach to find fermionic bound states must be contrasted to the results in which the scalar field
kink is considered as an external field as presented in \cite{yi} for the case of the $\vp^4$ theory and in \cite{brihaye}
for the case of the kink-antikink in the sine Gordon model. However, as we shall demonstrate
in the present work, the presence of the fermion changes
drastically the kink profile and the relevant spectrum.

The mass and interaction terms of the $\Phi^4$ theory can be considered as the first terms in
the expansion of the functional $U_{1}[\Phi]$ of our model. Therefore, the system (\ref{5211})-(\ref{phizetas1})
with $e_{\psi}=k/16$ and $\bar{\zeta} \equiv 0$, considering the mass and $\Phi^4$ interaction terms in the expansion of $U_1[\Phi]$, will be equivalent to the model recently presented in \cite{Shahkarami1}.

In this paper we provide some analytical solutions to the systems of eqs. (\ref{5211})-(\ref{phizetas1}). In order to tackle this problem we follow closely the tau function approach developed for the ATM model \cite{npb1}. This approach will be adapted conveniently for the static version of the ATM model (\ref{5211})-(\ref{phizetas1}) and implemented taking into account a variety of topologically nontrivial configurations for the scalar field $\Phi$ and the relevant fermionic bound states associated to the field $\xi_a$. We are interested in solutions for the field $\Phi$ representing a kink type solution such that
\br \Phi(x)&=&-\Phi(-x),\label{parity0}\\
 \Phi(\pm \infty)&=&\pm \phi_o,\,\,\,\,\Phi(0)=0,\,\,\,\,\phi_o= const. \label{infty0}.
\er

We consider those solutions to the system (\ref{5211})-(\ref{phizetas1}) in such a way that the Hamiltonian is
invariant under parity transformation, then the terms involving the auxiliary field $\bar{\zeta}$ and the functional $U_1[\Phi]$ must take convenient forms in order to leave invariant the  equation (\ref{phizetas1})  under parity transformation.

Under the above assumptions the solutions to the system (\ref{5211})-(\ref{phizetas1}) will be eigenfunctions of the parity operator. Then one must have
\br
\xi_{1}(x)= - \s\, \xi_{4}(-x),\,\,\,\, \xi_{2}(x) = \s\, \xi_{3}(-x),\,\,\,\,\,\s = \pm 1, \label{parity}
\er
where $\s$ is the parity operator eigenvalue. Moreover, one imposes the following conditions to the spinors
\br
\label{bcspinors}
\xi_{a}(\pm \infty) \rightarrow  0,\,\,\,a=1,...,4;\,\,\,\,\xi_{1}(0) =  \xi_{0}=const.
\er

An additional consistency condition must be satisfied by the solutions; in fact, from the equations (\ref{5211})-(\ref{5241}) one gets the relationship
\br
\label{consist}
\xi_{1}^2+ \xi_{2}^2-\xi_{3}^2-\xi_{4}^2 = c_1= const.
\er

Since all the spinor components vanish at $x\rightarrow \pm \infty$, according to the boundary condition (\ref{bcspinors}), from now on we will set $c_1 = 0$. The above set of conditions will restrict considerably the possible forms of the auxiliary field $\bar{\zeta}$ and the functional $U_1[\Phi]$ as discussed below. In addition to the above set of conditions we require a finite energy and localized energy density associated to the field $\Phi$ such that it becomes a  kink type solution of the system, so the functional $U_1[\Phi]$ in (\ref{phizetas1}) must have a degenerate vacua.

\section{Self-consistent solutions and the tau function approach}
\label{sec:tau}

Next, we solve the system of equations (\ref{5211})-(\ref{phizetas1}) analytically and determine the eigenvalue $E$. In order to do this we will follow the tau function approach and assume that the expressions of the fields $\xi_a$ and  $\Phi$ in terms of certain  tau functions to be closely related to the ones which have been proposed in the relativistic ATM model \cite{matter, npb1}. So, the pseudoscalar and the spinor fields of the model are assumed to depend on the tau functions $\{\tau_{1}^{(n)},\,\tau_{0}^{(n)},\, \tau_{\xi ,n}^{a},\,\widetilde{\tau}_{\xi ,n}^{a}\}$ by the Ansatz
\br
e^{i \Phi_n/n} &=& e^{-i \theta_n} \, \frac{\tau_{1}^{(n)}}{\tau_{0}^{(n)}},\,\,\,\,n \equiv 1,2,3,... \label{tau01}\\
\xi_{1, n} &=& \tau_{\xi ,n}^{1} \,[\frac{1}{\tau_0^{(n)}}]^n \, e^{-i \theta_{1,n}} +\widetilde{\tau}_{\xi ,n}^{1}\, [\frac{1}{\tau_1^{(n)}}]^n \, e^{i \theta_{1,n}} \label{tau1}\\
\xi_{2, n}&=& -i\{\tau_{\xi ,n}^{2}\, [\frac{1}{\tau_0^{(n)}}]^n \, e^{-i \theta_{2,n}} - \widetilde{\tau}_{\xi ,n}^{2}\, [\frac{1}{\tau_1^{(n)}}]^n  \, e^{i \theta_{2, n}}\} \label{tau2}\\
\xi_{3, n} &=& \tau_{\xi ,n}^{3}\, [\frac{1}{\tau_0^{(n)}}]^n \, e^{-i \theta_{3, n}} +\widetilde{\tau}_{\xi ,n}^{3} \, [\frac{1}{\tau_1^{(n)}}]^n \, e^{i \theta_{3, n}}\label{tau3} \\
\xi_{4, n} &=& i\{ \tau_{\xi ,n}^{4} \,[\frac{1}{\tau_0^{(n)}}]^n \, e^{-i \theta_{4,n}} -\widetilde{\tau}_{\xi ,n}^{4} \, [\frac{1}{\tau_1^{(n)}}]^n \, e^{i \theta_{4,n}}\},\label{tau4}
\er
where $\theta_n,\,\theta_{a,n}$ are real parameters. Notice that we have incorporated a positive integer $n\in \mathbb{Z}_{+}$ in the expressions for the fields, each $n$ labels a particular solution corresponding to a kink profile with specific nontrivial topological configuration $\Phi_n$ and a set of relevant spinor solutions $\xi_{a,\,n}$. 

In order to find a self-consistent solution to the  system of equations (\ref{5211})-(\ref{phizetas1}) we implement the procedure in two steps for each topological sector $n=1,2$. First, we substitute the Ansatz (\ref{tau01})-(\ref{tau4}) into the system (\ref{5211})-(\ref{5241}) for the spinors coupled to the field $\Phi_n$ and determine the eigenvalue $E_n$. Second, the eq. (\ref{phizetas1}) is solved by requiring a self-consistency condition in each topological sector. In fact, in the static version of the ATM system (\ref{5211})-(\ref{phizetas1}) the functionals  ${\bar{\zeta}}[\Phi],\,U_1[\Phi]$ associated to each topological sector $(n)$ must be chosen such that, for every set of solutions
$\{ \Phi_{n},\, \xi_{a, n}, \, (a=1,2,..4) \} $ satisfying the system (\ref{5211})-(\ref{5241}),
the kink $\Phi_{n}$ would remain as a self-consistent solution of the eq. (\ref{phizetas1}).

We will concentrate below on some properties in the cases $n=1,2$ by constructing the explicit solutions and computing explicitly some of the related quantities, such as the topological charges, bound state energies, scalar field potentials and their vacua, kink and boson masses, as well as the kink and the corresponding fermionic bound state profiles, respectively. In section \ref{sec:n3} we will provide a short discussion on the cases $n \geq 3$.

\subsection{The sector $n=1$ and the topological charge $Q_{topol} = \frac{\theta_o}{\pi}$}
\label{subsec:n1}

Let us set $n=1$ in (\ref{tau01})-(\ref{tau4}) and consider the tau functions
\br
\label{t01}
\tau_{0}^{(1)} &=& 1+ e^{- i \theta_1} e^{ 2 \kappa x },\,\,\,\tau_{1}^{(1)} = 1  + e^{ i \theta_1} e^{ 2 \kappa x },\\
\label{tauxi1}
\tau_{\xi ,1}^{1} &=& b_1 e^{ \kappa x },\,\, \tau_{\xi ,1}^{2}= b_2 e^{ \kappa x },\,\,\tau_{\xi ,1}^{3} = b_3 e^{ \kappa x },\,\, \tau_{\xi ,1}^{4}= b_4 e^{ \kappa x },\\\label{tauxi2}
\widetilde{\tau}_{\xi ,1}^{1} &=& \bar{b}_1 e^{ \kappa x },\,\, \widetilde{\tau}_{\xi ,1}^{2} = \bar{b}_2 e^{ \kappa x },\,\,\widetilde{\tau}_{\xi ,1}^{3} = \bar{b}_3 e^{ \kappa x },\,\,\widetilde{\tau}_{\xi ,1}^{4} = \bar{b}_4 e^{ \kappa x },\er
where $b_{a},\,\bar{b}_a$ are complex constants and $\kappa, \theta_1$ are real parameters. For simplicity let us choose $\theta_{a,1} \equiv \theta_1\,\, (a=1,2,...4)$.

The imposition of the parity conditions (\ref{parity}) to the $\xi_{a,1}$ fields defined in (\ref{tau1})-(\ref{tau4}) for the tau functions (\ref{t01})-(\ref{tauxi2})  provide the parameters relationships
\br
\label{barb1}
\bar{b}_3 = -i \s\, b_2 e^{-i \theta_n},\,\,b_3 = i \s\, \bar{b}_2 e^{i \theta_n},\,\, \bar{b}_4 = -i \s\, b_1 e^{-i \theta_n},\,\, b_4 = i \s\, \bar{b}_1 e^{i \theta_n}
\er

Consider the parametrization
 \br
 \label{parameters}
 b_2 = \rho_ 2 e^{i \a_2},\,\, \bar{b}_2 = \rho_ 2 e^{-i \alpha_2},\,\,  b_1 = \rho_1 e^{i \alpha_1},\,\, \bar{b}_1 = \rho_1 e^{-i \alpha_1}.
 \er

The spinor components become
\br
\label{spinors12}
\xi_{a,1} = (-i)^{a-1} \, \rho_a\, e^{[\kappa x -i (\a_a+\theta_1)]} \Big[ \frac{e^{(2 i \a_a)}}{1+e^{(2 \kappa x- i \theta_1)}}+ (-1)^{a-1} \frac{e^{(2 i \theta_1)}}{1+e^{(2 \kappa x+i \theta_1)}}\Big],\,\,\,\,a=1,2.
\er
The remaining components $\xi_{a,1}\,(a=3,4)$ can be obtained from the above expressions (\ref{spinors12}) taking into account the parity condition relationships (\ref{parity}). Notice that the expressions (\ref{spinors12}) define real  $\xi_{a,1}$ functions. Taking into account the relations (\ref{tau1})-(\ref{tau4}) and  the condition (\ref{consist}) the parameters  (\ref{parameters}), must satisfy
\br
\label{cond2}
\rho_1^2 \sin{(2 \alpha_1 - \theta_1)} =  \rho_2^2 \sin{(
    2 \alpha_2 - \theta_1)}.
\er

Then, the simplest choice $\a_1=\a_2,\,\,\rho_2=\pm \rho_1$ satisfies the above consistency condition (\ref{cond2}). On the other hand, the conditions (\ref{parity0})-(\ref{infty0}) applied to the function $\Phi_1$ as defined in (\ref{tau01}) for $n=1$  will restrict the possible values of the parameter $\theta_1$. So, the field $\Phi_1$ can be written as
\br
\label{kink1}
\Phi_1(x) = -\theta_1 + 2 \arctan{\Big\{\frac{\sin{(\theta_1)} \,\, e^{2 \kappa x}}{1+  \cos{(\theta_1)} \,\, e^{2 \kappa x} }\Big\} },\,\,\,\, \theta_1 \in I_{q} \equiv  [-\frac{\pi}{2}+ 2 \pi q\, ,\, \frac{\pi}{2}+ 2 \pi q],
\er
where $q \in \mathbb{Z}$. The asymptotic values of the field are $ \Phi_1 (\pm \infty)\equiv \pm \theta_1 \( \theta_1 \in  [-\frac{\pi}{2}\,,\, \frac{\pi}{2}]\)$. The case $n=2$ in (\ref{tau01}) will provide a kink with asymptotic values $\Phi_2(\pm \infty) \in [-\pi\,,\,\pi]$ studied in the literature, see below. The angle $\theta_1$ is restricted by the conditions (\ref{parity0}) and (\ref{infty0}) to belong to the intervals $I_{q}$ defined for any $q\in \mathbb{Z}$. The kink (\ref{kink1}) is plotted in Fig. 1. Define the topological charge as
\br
\label{topol1}
Q_{topol}^{(n)} &\equiv &\frac{1}{2 \pi}\,\int_{-\infty}^{+\infty} \pa_x \Phi_n\\
\label{topol11}
&=& \frac{\Phi_n(+\infty)-\Phi_n(-\infty)}{2 \pi}.
\er
In the case at hand it becomes  $Q_{topol}^{(1)}=\frac{\theta_1}{\pi}$, so it can take any non-integer value in the interval $Q_{topol}^{(1)} \in [-1/2\,,\,1/2]$.

The region of validity  of the angle $\theta_1$ must also be consistent with the relation (\ref{cond2}) and  the relationship between the parameters due to the  normalization condition on the fermion sector.

\begin{figure}
\centering
\includegraphics{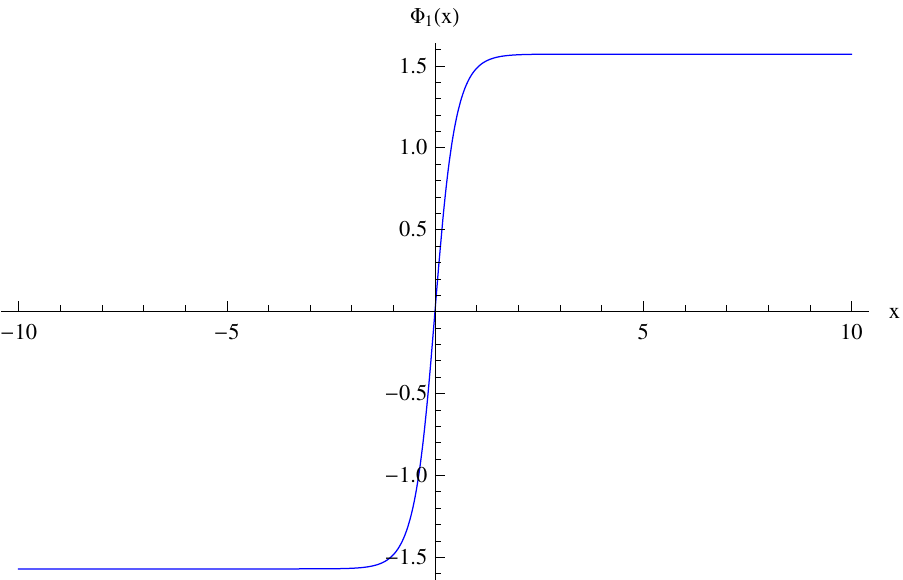}
\parbox{5in}{\caption{The  graph of the kink $\Phi_1(x)$ corresponding to the parameters
$\theta_1 = \pi/2,\, \kappa=1.57$. Notice that $\Phi_1(\pm \infty)=\pm \pi/2$ }}
\end{figure}

The boundary condition for the spinor $\xi_1$ is given by
\br
\label{cond1}
\xi_1(0)\equiv \xi_0 = \rho_1 \frac{\cos{(\frac{\theta_1}{2}-\a_1)}}{\sin{\frac{\theta_1}{2}}}
\er

The spinor is normalized as follows 
\br
\nonumber
\int_{-\infty}^{+\infty} \Big[ \sum_{a=1}^{4} \xi_{a}^2 \Big] dx &=& \frac{2}{|\kappa|} \times \\
\label{cond31}
&& \Big[ \theta_1 \csc{(
\theta_1 )} (\rho_1^2+\rho_2^2) + \rho_1^2 \cos{(2 \a_1 - \theta_1 )}-\rho_2^2 \cos{(2 \a_2 - \theta_1)}   \Big] \\
&=&1. \label{cond32}
\er

\begin{figure}
\centering
\includegraphics[width=5cm,scale=1, angle=0,height=5cm]{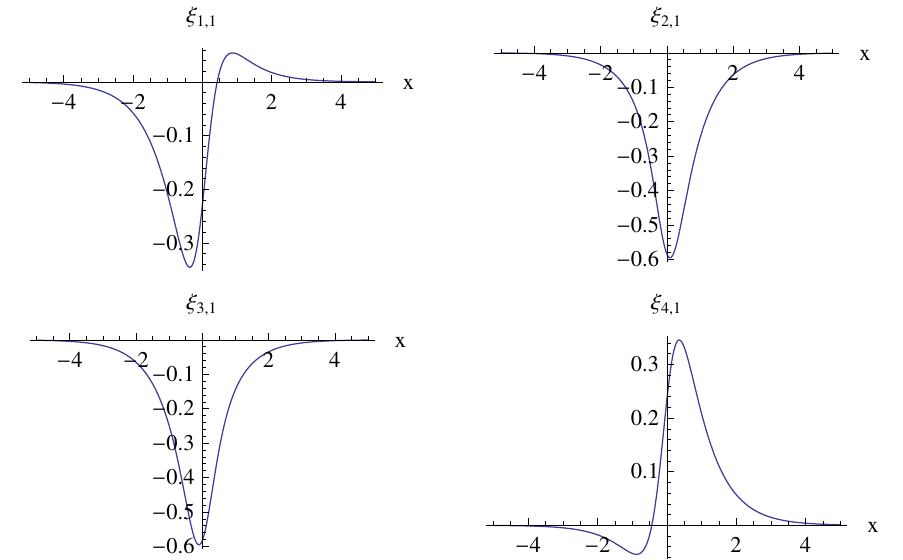}
\parbox{5in}{\caption{The bound state of the fermion as a function of $x$ for the {\bf positive parity} $\s=1$. It is plotted for $\rho_1 = \rho_2 = 0.5,\, \a_1 =\a_2 =-\theta_1 = -1.3,\, \kappa =1.35$. Notice the spinor components relationships $\xi_{1,1}(-x)= - \xi_{4,1}(x)$ and $\xi_{2,1}(-x)= + \xi_{3,1}(x)$}}
\end{figure}

\begin{figure}
\centering
\includegraphics[width=5cm,scale=1, angle=0,height=5cm]{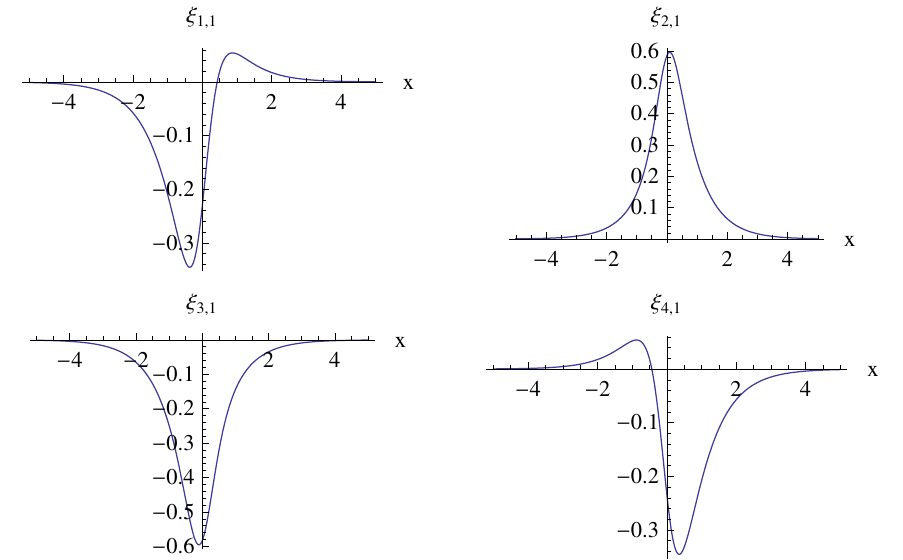}
\parbox{5in}{\caption{ The bound state of the fermion as a function of $x$ for the {\bf negative parity} $\s=-1$. It is plotted for $\rho_2 = -\rho_1 = -0.5,\, \a_1 =\a_2 =-\theta_1 = -1.3,\, \kappa =1.35$. Notice the spinor components relationships $\xi_{1,1}(-x)=+\xi_{4,1}(x)$ and $\xi_{2,1}(-x)= -\xi_{3,1}(x)$}}
\end{figure}

In the Figs. 2 and 3 we present the graphs of the solutions  $\xi_a,\,\,a=1,...4$ defined in (\ref{tau1})-(\ref{tau4}) considering the tau functions (\ref{t01})-(\ref{tauxi2}) for the both signs of the parity eigenvalues $\s= \pm 1$, respectively. The parameters in these Figs. satisfy the condition (\ref{cond2}) and the normalization condition (\ref{cond31})-(\ref{cond32}). One can see that the kink parameters $\kappa, \theta_1$ always depend on the spinor parameters through the eq. (\ref{cond2}), the boundary condition (\ref{cond1}) and the normalization condition (\ref{cond31})-(\ref{cond32}), therefore  the shape of the soliton always depends on the spinor parameters, and vice versa.

In order that the eq. (\ref{phizetas1}) to be satisfied one must choose the functionals $\bar{\zeta}$ and $U_{1}$ in a  consistent way. The substitution of the relationships (\ref{tau01})-(\ref{tau4}) given in terms of the tau functions (\ref{t01})-(\ref{tauxi2}) into the equation (\ref{phizetas1}) will determine the forms of $\bar{\zeta}$ and $U_1[\Phi_1]$. In fact, the eq.  (\ref{phizetas1}) is identically satisfied provided that
\br\nonumber
\bar{\zeta}^{''} & \equiv & 0 \\
U_1[\Phi_1 ] & \equiv & A_1 \cos{\Phi_1}+  A_2 \cos{(2 \Phi_1)}+ A_3 \cos{(3\Phi_1)},\label{pot1} \\
 A_1 &=& 3 R_1 \Big[ 4 \cos{\theta_1}(\frac{4 \s \kappa^2}{M_2} + B_1) + B_2\Big],\,\,\, A_2 = - 3 R_1 \Big[  \frac{4 \s \kappa^2}{M_2} + B_1 + B_2\Big],
\,\,\,\,A_3 = R_1  B_2,\,\,  \nonumber \\
B_1 &=& \rho_2^2 \cos{(2 \a_2-\theta_1 )}-\rho_1^2 \cos{(2 \a_1-\theta_1 )},\,B_2 = \rho_1^2 + \rho_2^2 - 2 \rho_1  \rho_2 \cos{(\a_1-\a_2)},\\
\,\,\,M_2 &\equiv & \frac{16 m_{\psi} e_{\psi} e^{\eta_0}}{k},\,\,\,R_1\equiv -\frac{\s \, M_2}{12 (\sin{\theta_1})^2}.\nonumber\er

Some comments are in order here. First, the functionals $\bar{\zeta}$ and  $U_1$ were chosen such that the eq. (\ref{phizetas1}) is identically satisfied, and their properties in general depend on the scalar $\Phi_1$, and consequently their behaviors under parity transformation will depend on the form of $\Phi_1$. Taking into account the relationships (\ref{parity0}) and (\ref{parity})  for the scalar and the spinors, respectively, under parity transformation, and  in order to maintain invariant the eq. (\ref{phizetas1}) under the parity transformation one has conveniently identified the vanishing sector with $\bar{\zeta}'' \equiv 0$ and the odd sector with the term involving the derivative $U_1'[\Phi_1]= \frac{\pa U_1}{\pa \Phi_1}$. In fact, $U_1[\Phi_1]$ is an even function under $\Phi_1 \rightarrow -\Phi_1$ as it is clear in (\ref{pot1}).

It is useful to consider the spinors $\xi_{a}[\Phi_1]$ as functionals of the scalar field $\Phi_1$. So, the terms involving the spinors in eq. (\ref{phizetas1}) will provide a functional $-U_2[\Phi_1]$. Then one is able to write an equation for the scalar
\br
\label{phiestatic}
\Phi_1^{''} = \frac{\pa U}{\pa \Phi_1},\,\,\,\,\,\,\,U \equiv U_1[\Phi_1] + U_2[\Phi_1]
\er
with
\br
U[\Phi_1]&=& \frac{\kappa^2}{(\sin{\theta_1})^2} [ - 4 \cos{\theta_1} \cos{\Phi_1} + \cos{(2 \Phi_1)}]+ const. \label{potential}\\
\nonumber
U_2[\Phi_1]&=&\frac{M_2 \s}{6 (\sin{\theta_1})^2} \cos{\Phi_1} \{ \rho_1^2 + \rho_2^2 + 6 B_1 \cos{\theta_1} - 6 [\rho_2^2 \cos{\a_2} \cos{(\a_2-\theta_1)} +\\ && 
\nonumber \rho_1^2 \sin{\a_1} \sin{(\a_1-\theta_1)} -  \rho_1 \rho_2  \cos{(\a_1-\a_2)}  \cos{\theta_1} ]  \cos{\Phi} -\nonumber \\
&& 4 \rho_1 \rho_2  \cos{(\a_1-\a_2)} (\cos{\Phi_1})^2 + (\rho_1^2 + \rho_2^2)  \cos{2 \Phi_1} \} .
\er

Second, setting to zero the derivative of the potential (\ref{potential}) 
\br
U'[\Phi_1] \sim (\cos{\theta_1} - \cos{\Phi_1}) \sin{\Phi_1}= 0
\er
one gets the vacuum points  $\Phi_{1,vac} = \{ \theta_1 + 2 \pi n_1,\,\,\pi n_2\,\, / \,n_1, \, n_2 \in \mathbb{Z}\}$. The vacuum  $\Phi_{1, vac} =\theta_1$ was expected since it corresponds to the kink in (\ref{kink1}). In addition, the vacua $\Phi_{1, vac}= \pi n_2 $ are inherited from the original ATM model in (\ref{atm1}). A representative soliton interpolating these type of vacua is presented in (\ref{sol1}).

For the boson mass one gets
\br\nonumber
m^2 & \equiv &- \frac{\pa^2 U[\Phi_1]}{\pa \Phi_1^2} |_{\Phi_1=0}\\
 &=&  2 \kappa^2 (\sec{\frac{\theta_1}{2}})^2.\label{mass1}
\er

The soliton mass associated to the field $\Phi_1$ given in (\ref{kink1}) is obtained from the relation $M_{sol} =  \int_{-\infty}^{+\infty} (\Phi_1^{'})^2 $, so one has
\br
 M_{sol} = 4 |\kappa|\, [ 1 -\theta_1 \cot{(\theta_1)} ]. \label{solmass1}
\er

It is a remarkable fact that the energy density associated to the scalar in (\ref{phiestatic}) \cite{rajaraman}
\br
{\cal \epsilon}(x) = \frac{1}{2} (\Phi')^2 +  U[\Phi_1]= 2U[\Phi_1],\er
provides the soliton mass (\ref{solmass1}) for the kink (\ref{kink1}). The soliton mass parameters $\kappa, \, \theta_1$ will be related later to the spinor parameters through (\ref{cond2}) and the spinor normalization condition  (see below).

The changes in the local features of the soliton, once the asymptotic values $\Phi_1(\pm \infty)$ are fixed, will be considered below to study the spectral flow, so for later convenience we compute the slope of the kink at the origin $x=0$ 
\br
\label{slope1}
\mu_1 = 2 \kappa\, \tan{(\theta_1/2)}
\er

Next, we consider the problem of finding the bound state energy $E_1$. Substituting the relationships (\ref{tau01})-(\ref{tau4}) given in terms of the tau functions (\ref{t01})-(\ref{tauxi2}) into the equations (\ref{5211})-(\ref{5241}) one gets  a homogeneous linear system of equations for the unknown independent parameters $\{ b_1, b_2\}$.  Then imposing the condition of vanishing determinant for the $2 \times 2$ matrix formed by the coefficients in that linear system of equations, in order to find a non-trivial solution, one gets a quadratic algebraic equation for the bound state energy $E_1$   
 \br
 \nonumber
\Big[ 8 \kappa^2 &-& M_1^2+ M_1^2 \cos{(2 \theta_1)}\Big] E_1^2+ 2  \kappa M_1^2 \sin{(2 \theta_1 )}  E_1 + \kappa^2 M_1^2 - M_1^4-\kappa^2 M_1^2 \cos{(2 \theta_1)} + \\ 
&& M_1^4 \cos{(2 \theta_1)}  - 2 \sigma\, M_1 \{[ 4  \kappa^2 -  M_1^2 +  M_1^2 \cos{(2 \theta_1)} ] E_1 + \kappa M_1^2 \sin{(2 \theta_1)} \} = 0 \label{quadratic1}
 \er
where $M_1 = e_{\psi} m_{\psi} e^{\eta_o}$. In Figs. 5, 6 we provide some contour-plots for $E_1 \,\mbox{vs}\, \theta_1$ and in the Figs. 7 and 8  we plot $E_1 \,\mbox{vs}\, \rho_1$ for a special set of parameters, see below more on this point. Notice that $E_1(\theta_1=0) = \frac{M_1}{2} (\s + \epsilon_0),\,\,\epsilon_0=\pm 1$ . So,  $E_1(\theta_1=0) = \s M_1$, i.e., there is a positive-parity threshold bound state at $E_1 = M_1$ and a negative-parity threshold
bound state at $E_1 = - M_1$ (see the bottom right graph in Fig. 6).  

In the infinite limit of the slope $\mu \rightarrow +\infty$ ($\kappa \rightarrow +\infty$) one has
\br
\label{energyslopeinfty}
E_1(\theta_1) =\frac{1}{2} \s \, M_1 \( 1 \pm | \cos{\theta_1}|\)
\er
whereas, in the limit $\mu \rightarrow 0$  ($\kappa \rightarrow 0$) one gets
\br
E_1(\theta_1) = \s M_1
\label{energyslopezero}
\er

The result (\ref{energyslopeinfty}) must be contrasted with the exact results in  \cite{gousheh, mackenzie} for the piecewise linear potential $\Phi$  with a constant slope in the middle region. In this limit the authors have obtained $E_1 =M_1 \cos{\theta_o}$\,\,($0\leq \theta_o \leq \pi$). However, in those references the shape of the potential $\Phi$ is prescribed. 

Let us investigate the boson mass dependence of the kink energy. Consider the expression (\ref{solmass1}) of the soliton mass and the boson mass (\ref{mass1}), then one has
\br
\label{msolm}
M_{sol} &=& \{ 2 \sqrt{2}  [ 1- \theta_{1} \cot{\theta_{1}}] \cos{(\frac{\theta_{1}}{2})}\}\,\,m.
\er
This is an exact result and shows that the soliton mass is directly proportional to the boson mass. However, its expansion in powers of $\theta_{1}$ becomes $M_{sol} \approx \sqrt{2}\, m\, \theta_{1}^2 (1- \theta_{1}^2/8 + ...)$. This can be compared to the $\Phi^4$ theory exact result $M_{kink}^{\Phi^4} = \frac{2 \sqrt{2}}{3}\, m \, \theta_{1}^2$.

The boson mass dependence of the bound state energy can be obtained substituting  the relationship (\ref{mass1}) for $\kappa$ into the eq. (\ref{quadratic1}). In fact, this procedure provides an implicit eq. for $E$ and $m$. In Fig. 10 it is plotted $E$ vs $m$ for $M_1=1$.

On the other hand, one has the ratio $M_{sol}/m =2$ for  $\theta_1=\pi/2$. This ratio will be discussed below in connection to the Skyrme conjecture. In the Fig. 22 this ratio has been plotted in terms of $\theta_1$ (see the curve labeled by $n=1$).

\subsubsection{Special set of parameters}

Next we discuss a special set of parameters and the fermionic bound state energies in the case $n=1$. For simplicity in what follows we set $\a_2=\a_1,\,\rho_2=\rho_1$. This choice  satisfies identically the relationship (\ref{cond2}), and the normalization condition (\ref{cond31}) becomes
\br
\label{cond5}
\int_{-\infty}^{+\infty} \Big[ \sum_{a=1}^{4} \xi_{a}^2 \Big] dx = \frac{4 \rho_1^2}{|\kappa|}  \theta_1 \csc{(
\theta_1)}
\er
The fermion wave function normalization to unity provides a further restriction on the parameters, so
\br
\label{cond6}
\frac{4\rho_1^2}{|\kappa|} = \frac{\sin{\theta_1}}{\theta_1},\,\,\,\theta_1=[-\pi/2\,,\,\pi/2].
\er

\begin{figure}
\centering
\includegraphics[width=8cm,scale=1, angle=0,height=8cm]{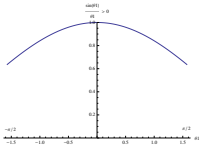}
\parbox{3in}{\caption{The graph of the function $\frac{\sin{\theta_1}}{\theta_1}>0$ for $\theta_1 \in [-\pi/2\,,\,\pi/2]$ which must be equal to the ratio $\frac{4 \rho_1^2}{|\kappa|}$ as in (\ref{cond6}).}}
\end{figure}

In the Fig. 4 it is plotted the positive region of the
function $ [\frac{\sin{\theta_1}}{\theta_1}] (> 0)$ which appears in (\ref{cond6}) defining the values of  $\kappa$ once the parameter $\rho_1$ is fixed.

A general region of validity of $\theta_1$ must  be consistent with the intervals $I_0$ in (\ref{kink1}), the relationship (\ref{cond2}) and the normalization condition (\ref{cond31})-(\ref{cond32})  determined for each choice of the set $\{\a_2, \a_1,\rho_2, \rho_1, \kappa \}$.  This amounts to generalize the simplest case (\ref{cond5})-(\ref{cond6}) considered above and we will not pursue it further.

\begin{figure}
\centering
\includegraphics[width=10cm,scale=1, angle=0,height=10cm]{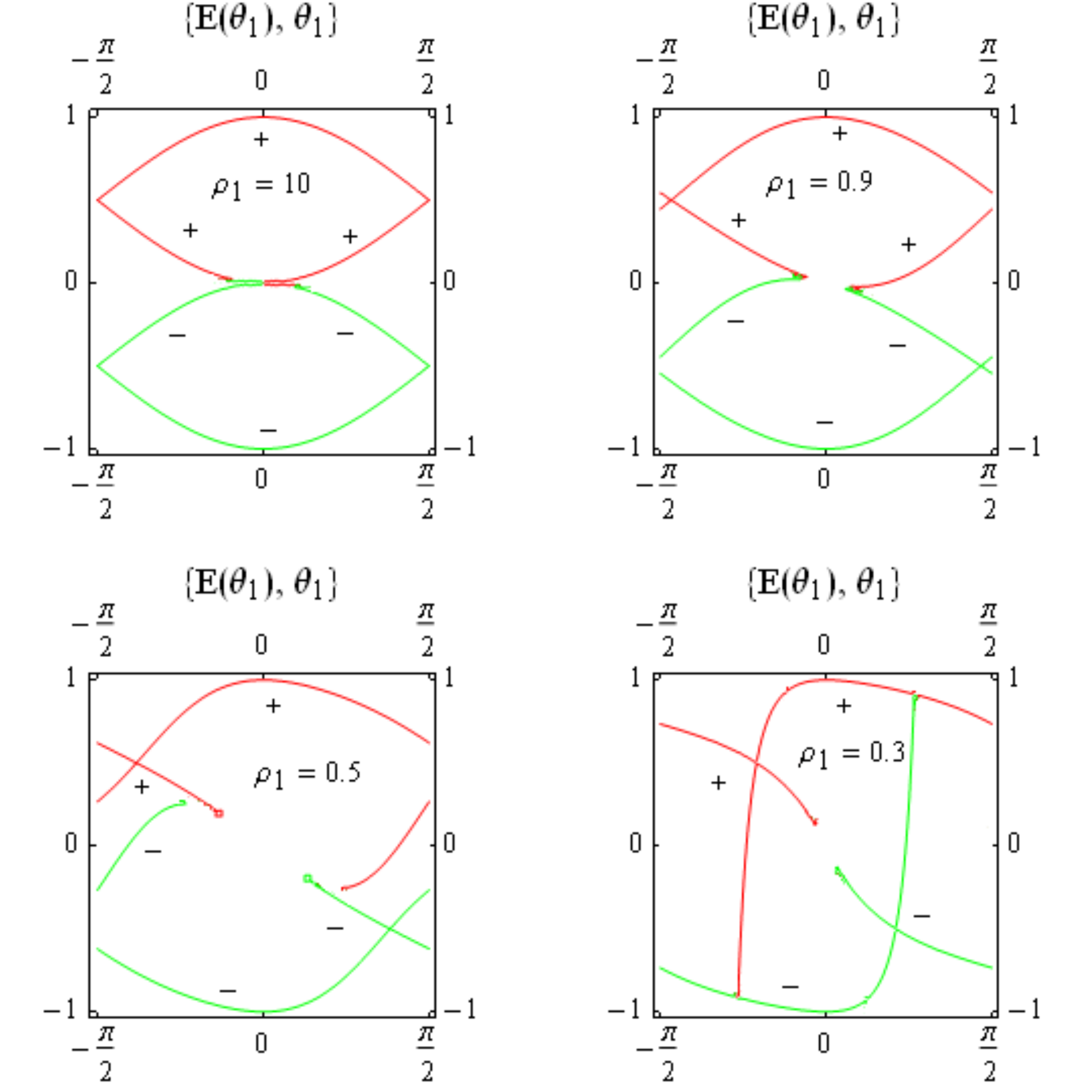}
\parbox{5in}{\caption{The fermionic bound state energies when $M_1=1$ and the values
$\rho_1=10,0.9,0.5,0.3$. The signs $\pm$ indicate the parity of the bound states. Notice
that for $\rho_{1}$ ( or $\kappa) \rightarrow + \infty $ , the graphs look like the function
 (\ref{energyslopeinfty})  for the both parities.}}
\end{figure}

\begin{figure}
\centering
\includegraphics[width=10cm,scale=1, angle=0,height=9cm]{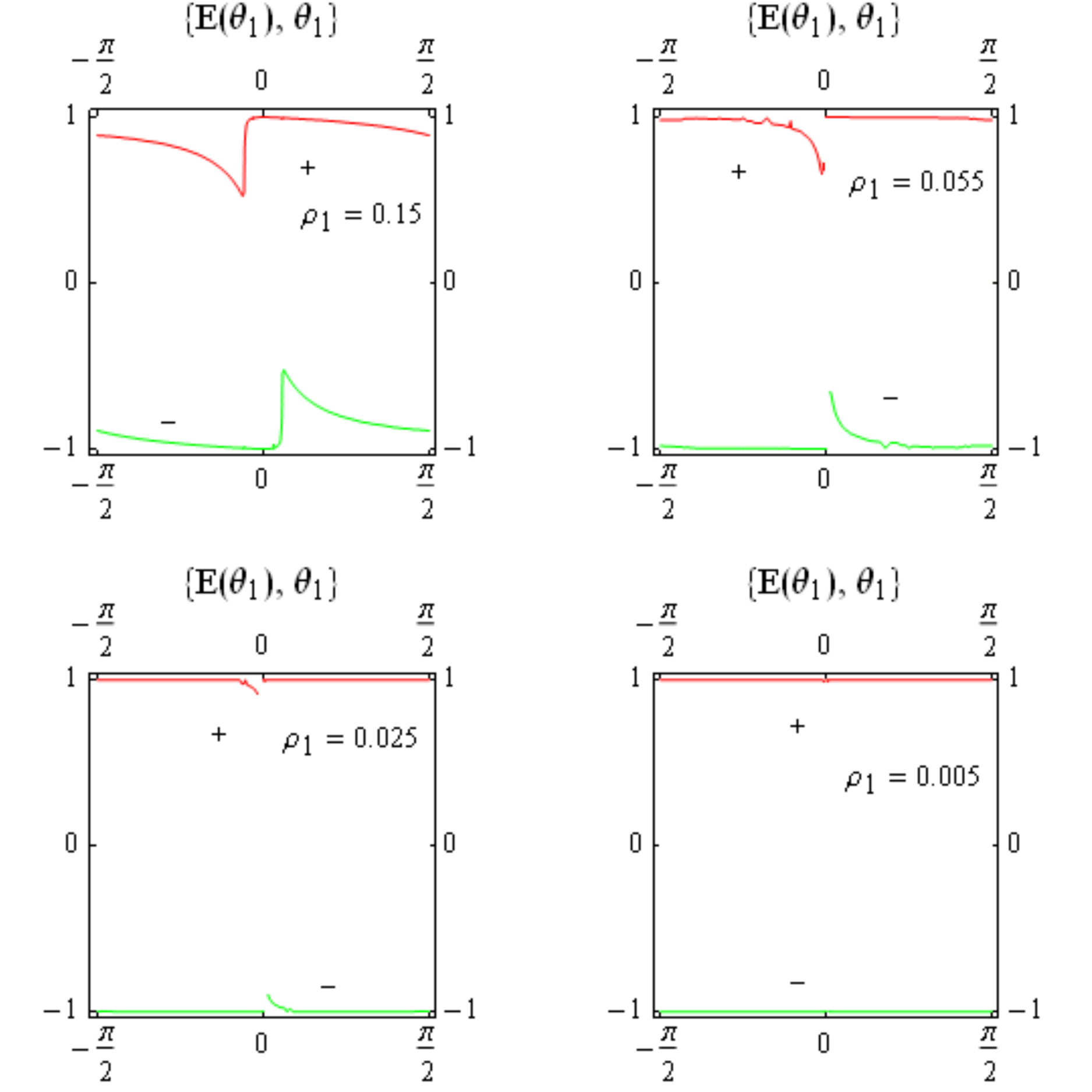}
\parbox{5in}{\caption{The fermionic bound state energies when $M_1=1$ and the values  $\rho_1=0.15, 0.055, 0.025, 0.005$.
The signs $\pm$ indicate the parity of the bound states.
Notice that for $\rho_{1}$ (or $\kappa) \rightarrow $ 0,
they reproduces (\ref{energyslopezero}), for the both parities.}}
\end{figure}

\begin{figure}
\centering
\includegraphics[width=10cm,scale=1, angle=0,height=8cm]{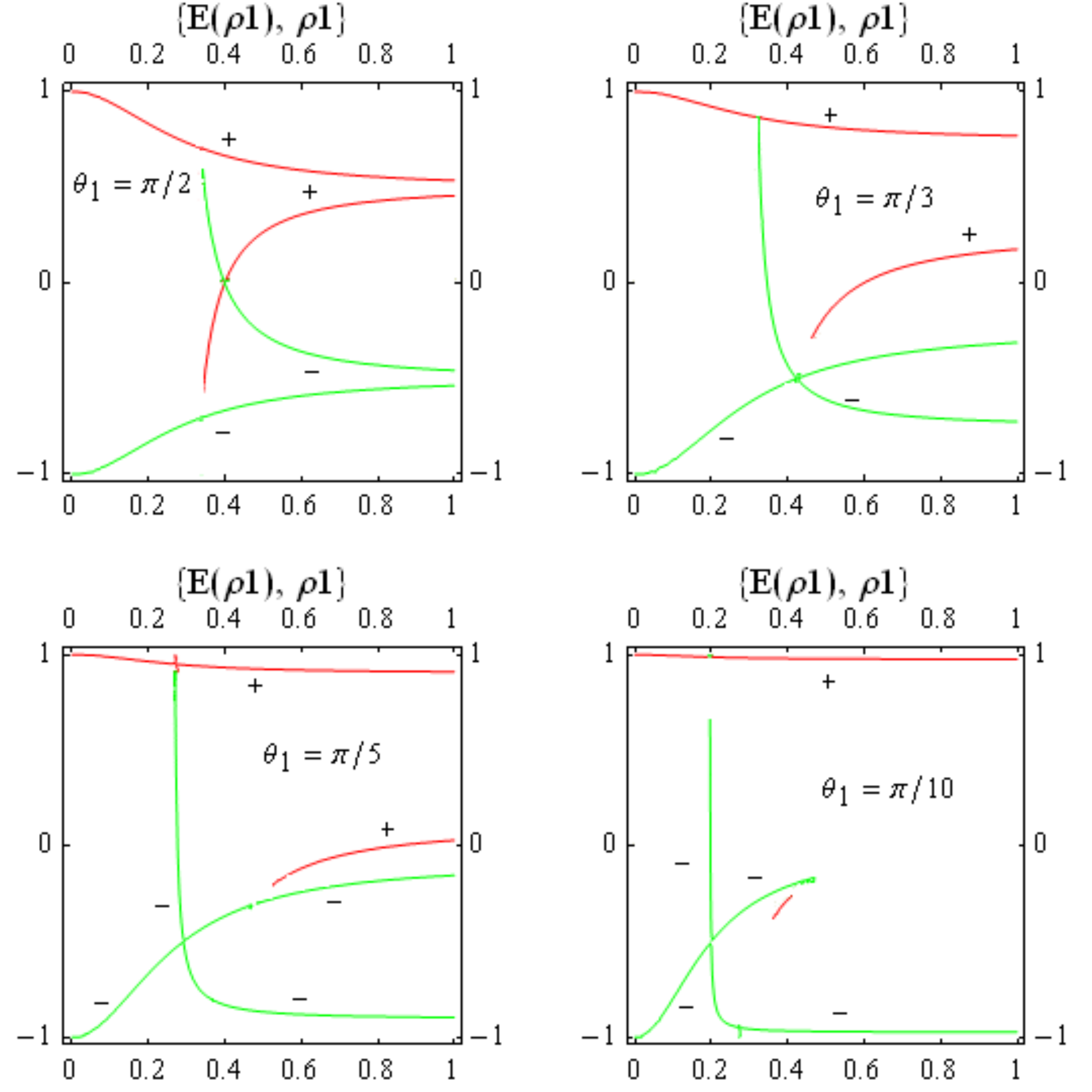}
\parbox{5in}{\caption{The fermionic bound state energies when $M_1=1$ and the values  $\theta_1= \pi/2, \pi/3,\pi/5,\pi/10$.
The signs $\pm$ indicate the parity of the bound states. 
Notice that for $\theta_{1}=\pi/2$ and $\rho_1 > 0.35$ there are four bound states, two of them for each parity; whereas for  $\rho_1 < 0.35$  there appears just one bound state for each parity. Notice that as $\theta_1$ decreases the number of positive parity bound states decreases.}}
\end{figure}

\begin{figure}
\centering
\includegraphics[width=10cm,scale=1, angle=0,height=8cm]{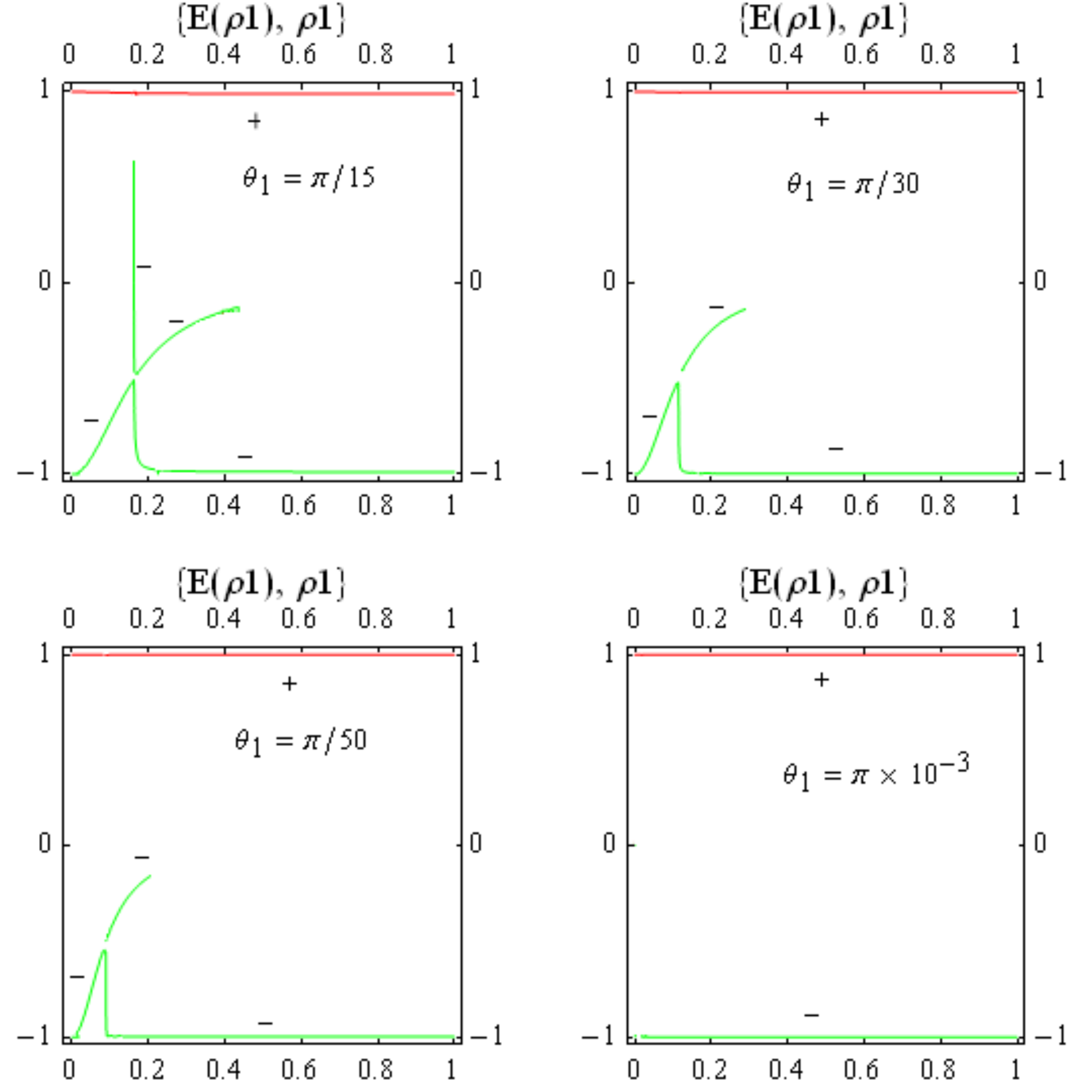}
\parbox{5in}{\caption{The fermionic bound state energies when $M_1=1$ and the values  $\theta_1=\pi/15,\pi/30,\pi/50, \pi \times 10^{-3}$.
The signs $\pm$ indicate the parity of the bound states. For $\theta_{1} \rightarrow 0$  the bottom right plot reproduces (\ref{energyslopezero}), for the both parities.}}
\end{figure}

Some comments are in order regarding the Figs. 5 and  6 corresponding to the energy spectrum of the $n=1$ sector.
First, the graphs of the bound state energies for $\theta_1$ negative are the reflection of the positive part
 through the origin $(0,0)$ provided the parity labels are changed. Second,
for $\theta_1$ positive, the number of bound states
and their separation from the positive (negative) continuum increases as the parameter $\rho_1$ is increased. Third, for some regions in the parameter space  there are up to four bound states, two for each parity. 

In order to investigate the spectral flow dependence on the local features of the soliton, once the asymptotic values $\Phi(\pm \infty)$ are fixed for a particular value of $\theta_1$, we have plotted $E$ vs $\rho_1$ in Figs. 7 and 8. For example, one observes that the energy level crosses $E=0$ for $\theta_1=\pi/2$ at $\rho_1=0.4$ as shown in the top left of Fig. 7. Notice that in the limit $\theta_1 \rightarrow  0$ one gets the Dirac sea corresponding to a trivial kink configuration as shown in the bottom right of Fig. 8. Moreover, from the Figs. 7 and 8 one notices that as $\theta_1$ decreases the positive parity bound states approach the positive continuum faster than the ones with negative parity do the negative continuum.

\begin{figure}
\centering
\includegraphics[width=5cm,scale=1, angle=0,height=5cm]{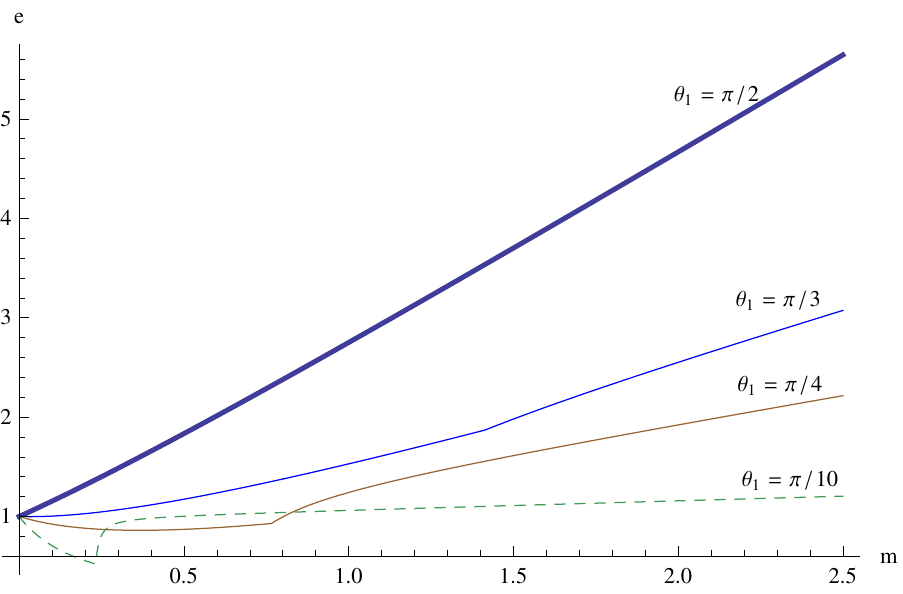}
\parbox{5in}{\caption{The total energy $e_1$ (soliton mass plus the fermionic bound sate energy with $\s=+1$)
as a function of the boson mass for $M_1=1, \theta_1 = \pi/2,\,\pi/3,\,\pi/4,\,\pi/10$. Notice the linear
behavior in the case $\pi/2$. So, it reproduces the result reported in \cite{Shahkarami1} through a
numerical approach. Notice that for a  wide range of values of the parameters $\theta_1$ and $m$ one has $e_{1} > M_1$ and the linear behavior of $e_1$.}}
\end{figure}

The Fig. 9 shows the total energy (soliton mass plus the fermionic bound sate energy with $\s=+1$) as a function of the boson mass for $M_1=1$ and some values of the angle $\theta_1$. Notice that for $\theta_1 = \pi/2$ one has a linear behavior resembling the result of ref. \cite{Shahkarami1}.

\begin{figure}
\centering
\includegraphics[width=5cm,scale=1, angle=0,height=5cm]{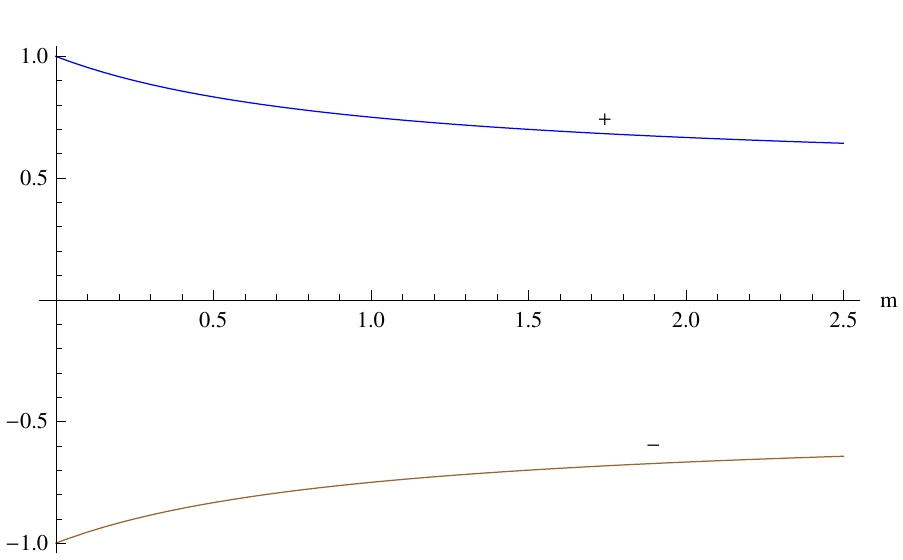}
\parbox{5in}{\caption{ The bound state energy of the fermion as a function of the boson mass $m$ for $M_1 = 1$ and $\theta_1= \pi/2$. The $\pm$ signs indicate the corresponding parity. Notice that the curve $E_1(m)$ never approaches zero as $m$ increases.}}
\end{figure}

\subsection{The sector $n=2$ and the topological charge $Q_{topol}=\frac{2 \theta_o}{\pi}$}
\label{subsec:n2}

Here we  closely follow the relevant steps of the previous construction. Set $n=2$ and $\theta_{a,2}\equiv \theta_2\, (a=1,2,...4) $ in (\ref{tau01})-(\ref{tau4}) and consider the tau functions
\br
\label{t01c}
\tau_{0}^{(2)} &=& 1+ e^{- i \theta_2} e^{ 2 \kappa x },\,\,\,\tau_{1}^{(2)} = 1  + e^{ i \theta_2} e^{ 2 \kappa x },\\
\nonumber
\tau_{\xi ,2}^{1} &=& b_1 e^{ \kappa x }+b_{11} e^{ 3 \kappa x },\,\, \tau_{\xi ,2}^{2}= b_2 e^{ \kappa x }+b_{22} e^{ 3 \kappa x },\\
\label{tauxi1c} \tau_{\xi ,2}^{3} &=& b_3 e^{ \kappa x }+b_{33} e^{ 3 \kappa x },
\,\, \tau_{\xi ,2}^{4}= b_4 e^{ \kappa x }+b_{44} e^{ 3 \kappa x },\\
\nonumber
\widetilde{\tau}_{\xi ,2}^{1} &=& \bar{b}_1 e^{ \kappa x }+\bar{b}_{11} e^{ 3\kappa x },\,\, \widetilde{\tau}_{\xi ,2}^{2} = \bar{b}_2 e^{ \kappa x }+\bar{b}_{22} e^{ 3\kappa x },\\ \widetilde{\tau}_{\xi ,2}^{3} &=& \bar{b}_3 e^{ \kappa x }+\bar{b}_{33} e^{ 3\kappa x },\,\,\widetilde{\tau}_{\xi ,2}^{4} = \bar{b}_4 e^{ \kappa x }+\bar{b}_{44} e^{ 3\kappa x },\label{tauxi2c}\er
where $b_{a},\,\bar{b}_a,\,b_{aa},\,\bar{b}_{aa},\,a=1,2,3,4$ are complex constants and $\kappa, \theta_2$ are real parameters.  Then the parity condition (\ref{parity}) provides the parameters relationships
\br
\label{relations11}
\bar{b}_1 &=& -i \s\, b_{44}  ,\,\,\bar{b}_{11}= -i \s\, b_4  ,\,\, \bar{b}_2 = -i \s\, b_{33}  ,\,\, \bar{b}_{22} = -i \s\, b_3,\, \bar{b}_3 = -i \s\, b_{22}  ,\\ \bar{b}_{33} &=& -i \s\, b_2  ,\,\, \bar{b}_4 = -i \s\, b_{11}  ,\,\, \bar{b}_{44} = -i \s\, b_1.\label{relations12}
\er
Let us choose the next parametrization which is consistent with the relationships (\ref{relations11})-(\ref{relations12})
 \br
 \label{parametersc1}
 b_a &=& \rho_ a e^{i \a_a},\,\, \bar{b}_a = \rho_a e^{-i \alpha_a},\\
 b_{aa} &=& \eta_a e^{i \b_a},\, \bar{b}_{aa} = \eta_a e^{-i \b_a},\,\,\,a=1,2,...4.\label{parametersc2}\\
 \b_a &=&- \a_a + \pi/2,\,\,\eta_1=\s \rho_4,\,\eta_2=\s \rho_3,\,\eta_3=\s \rho_2,\,\eta_4=\s \rho_1.\label{parametersc3}
 \er
 
Then, taking into account the relations (\ref{tau1})-(\ref{tau4}) and  the condition (\ref{consist}) the parameters  (\ref{parametersc1})-(\ref{parametersc3}), for $c_1=0$, must satisfy
\br
\nonumber
\rho_1^2 \sin{(2 \alpha_1)}-\rho_2^2 \sin{(2 \alpha_2)}-\rho_3^2 \sin{(2 \alpha_3)}+\rho_4^2 \sin{(2 \alpha_4)} &-&
 2 \s \,\Big[ \rho_2 \rho_3 \cos{(
     \alpha_2 - \a_3)} - \\ && \rho_1 \rho_4 \cos{(
     \alpha_1 - \a_4)}\Big]=0 \label{cond2c}
\er

From (\ref{tau1})-(\ref{tau4}) and (\ref{t01c})-(\ref{tauxi2c}), and considering the above parameterizations the spinors become
\br
\nonumber
\xi_{a,2}&=& [-i]^{(a-1)} \rho_a \exp{(\kappa x)} \Big\{ \frac{e^{i (\a_a-\theta_2)} + i \s \G_a\, e^{[-i (\g_a +\theta_2)]}  \exp{(2 \kappa x)}}{[1+ \exp{(2 \kappa x -i \theta_2)}]^2} + \\
&& (-1)^{a-1} \frac{e^{i (-\a_a+\theta_2)} - i \s \G_a\, e^{[i (\g_a +\theta_2)]}  \exp{(2 \kappa x)}}{[1+ \exp{(2 \kappa x + i \theta_2)}]^2} \Big\},\,\,\,\ a=1,2\label{spinorsn2}
\er
where $\G_1 \equiv  \rho_4/\rho_1,\,\G_2 \equiv \rho_3/\rho_2$,\,$\g_1 \equiv \a_4,\,\g_2 \equiv\a_3$. The expressions  (\ref{spinorsn2}) define real functions. The remaining components $\xi_{a,2}\,(a=3,4)$ can be obtained from the above expressions (\ref{spinorsn2}) taking into account the parity condition relationships (\ref{parity}).

From (\ref{tau01}) for $n=2$ and (\ref{t01c}) the field $\Phi_{2}$ can be written as
\br
\label{kink1c}
\Phi_2(x) = -2 \theta_2 + 4 \arctan{\Big\{\frac{\sin{(\theta_2)} \,\, e^{2 \kappa x}}{1+  \cos{(\theta_2)} \,\, e^{2 \kappa x} }\Big\} } ,\,\,\,\, \theta_2 \in I_{q} \equiv  [-\frac{\pi}{2}+ 2 \pi q\, ,\, \frac{\pi}{2}+ 2 \pi q], 
\er
where $q \in \mathbb{Z}$. Notice that the asymptotic values of the field
$ \Phi_2 (\pm \infty)\equiv \pm 2 \theta_2 \( \theta_2 \in  [-\frac{\pi}{2}\,,\, \frac{\pi}{2}]\)$.
The angle $\theta_2$ is restricted by the conditions (\ref{parity0}) and (\ref{infty0}) to
belong to the intervals $I_{q}$ defined for any $q\in \mathbb{Z}$. This kink is qualitatively
similar to the kink $\Phi_1$, however the kink in (\ref{kink1c}) approaches the asymptotic values
 $(\pm 2 \theta_2) \in [-\pi\,,\,\pi]$. These kinks are plotted in Fig. 11. So,
according to the definition (\ref{topol1}) the corresponding  topological charge becomes
$Q_{top}^{(2)} = 2 \theta_2/\pi$, which can take any non-integer value in the interval $\in <-1\,,\,1>$ and the integer $\pm 1$ charges just for $\theta_2= \pm \pi/2$, respectively. The slope of the kink at the origin $x=0$ becomes
\br
\label{slope2}
\mu_2 = 4 \kappa\, \tan{(\theta_2/2)}
\er

\begin{figure}
\centering
\includegraphics[width=5cm,scale=1, angle=0,height=5cm]{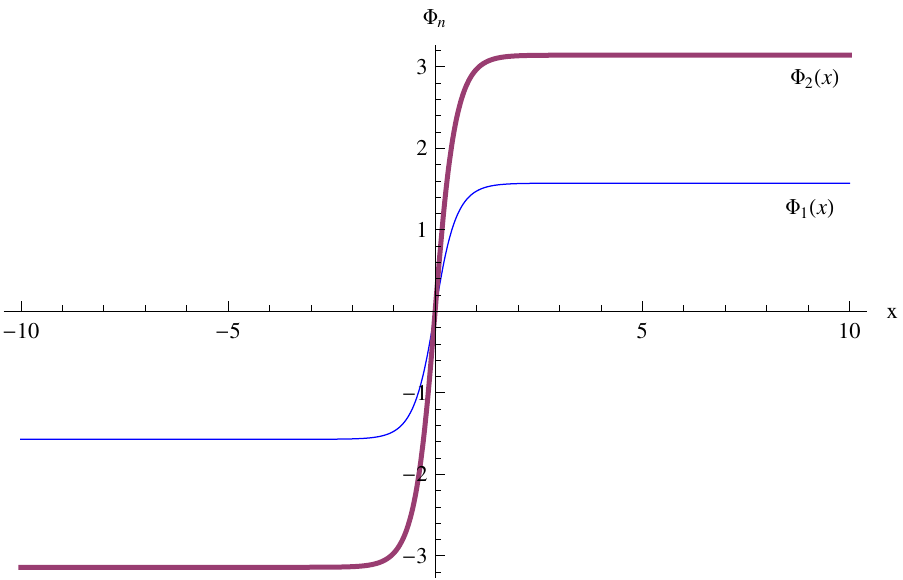}
\parbox{5in}{\caption{The  graphs of the kinks $\Phi_n(x)\,(n=1,2)$ corresponding to the parameters
$\theta_1 = \theta_2=\pi/2,\, \kappa_1 =1.35,\,\kappa_2=1.57\, (\rho_o =0.5)$. Notice that $\Phi_1(\pm \infty)=\pm \pi/2,\,\Phi_2(\pm \infty)= \pm \pi.$ At $x=0$ the slope of the kink $2$ is greater than the one of the kink $1$ for this choice of parameters.}}
\end{figure}

The region of validity $I_q$ of the angle $\theta_2$ would be further restricted when the condition (\ref{cond2c}) on the parameters and  the normalization condition on the fermion sector are imposed below.

The soliton mass associated to the field $\Phi_2$ given in (\ref{kink1c}) becomes
\br
 M_{sol} = 16 |\kappa|\, [ 1 -\theta_2 \cot{(\theta_2)} ]. \label{solmass1c}
\er
The potential associated to the scalar field $\Phi_2$ will be discussed below.

\subsubsection{Special set of parameters}

Following the same steps as in the case $n=1$, one has that the eq. (\ref{phizetas1}) is satisfied provided that the functionals $\bar{\zeta}[\Phi_2]$ and $U_{1}^{(2)}[\Phi_2]$ are chosen in a  consistent way. The substitution of the relationships (\ref{tau01})-(\ref{tau4}) for $n=2$, given in terms of the tau functions (\ref{t01c})-(\ref{tauxi2c}) into the equation (\ref{phizetas1}) will determine the expressions for $\bar{\zeta}$ and $U_1^{(2)}[\Phi]$.
For simplicity we perform this process for a special set of parameters $\rho_a=\rho_o,\,\,\a_a=\a_o,\,\,a=1,2,...4$. So, one can identify the vanishing sector with $\bar{\zeta}'' \equiv 0$ and the odd sector with $U_1^{(2)\, '}[\Phi_2] = \frac{\partial U_1^{(2)}}{\partial \Phi_2} $ for the functional $U_1^{(2)}[\Phi_2]$ given by
\br
\label{u1n2}
U_1^{(2)}[\Phi_2] = 2 (\csc{\theta_2})^3 [ M_2 \rho_o^2 \cos{(2 \alpha_o)} +
   4 \kappa^2 \sin{\theta_2}] \cos{(\frac{\Phi_2}{2})}\, [2 \cos{\theta_2} - \cos{(\frac{\Phi_2}{2})}].
\er

The expression (\ref{u1n2}) has been obtained by setting $\rho_a=\rho_o,\,\a_{a} =\a_{o}$ in the eq. (\ref{u1n22}). In fact, $U_1^{(2)}[\Phi_2]$ is an even function under $\Phi_2 \rightarrow -\Phi_2$. Next, considering the spinors $\xi_{a}[\Phi_2]$ as functionals of the scalar field $\Phi_2$, the terms involving the spinors in eq. (\ref{phizetas1}) will provide $\{ -U_2^{(2)\, '}[\Phi_2] \}$ associated to the functional
\br
U_2^{(2)}[\Phi_2] =- 2 (\csc{\theta_2})^3 [ M_2 \rho_o^2 \cos{(2 \alpha_o)} +
   8 \kappa^2 \sin{\theta_2}] \cos{(\frac{\Phi_2}{2})}\, [2 \cos{\theta_2} - \cos{(\frac{\Phi_2}{2})}]\label{u2n2}
\er
where the eq. (\ref{u2n22}) has been used for $\rho_a=\rho_o,\,\a_{a} =\a_{o}$.  Then one is able to write an equation for the scalar
\br
\label{phiestaticc}
\Phi''_2 = \frac{\pa U^{(2)}}{\pa \Phi_2},\,\,\,\,\,\,\,U^{(2)} \equiv U_1^{(2)}[\Phi_2] + U_2^{(2)}[\Phi_2]
\er
with
\br
U^{(2)}[\Phi]&=& \frac{4 \kappa^2}{(\sin{\theta_2})^2} [ - 4 \cos{\theta_2} \cos{(\Phi_2/2)} + \cos{\Phi_2}]+ const. \label{potentialc}
\er

Setting to zero the derivative of the potential (\ref{potentialc}) 
\br U^{(2)\, '}[\Phi] = \frac{8 \kappa^2}{(\sin{\theta_2})^2}  [ \cos{\theta_2} - \cos{(\frac{\Phi_2}{2})}] \sin{(\frac{\Phi_2}{2})},\er
 one gets the vacuum points $\Phi_{vac} = \{ 2 \theta_2 + 2 \pi n_1,\,\, 2 \pi n_2 \,\, / \,n_1, \, n_2 \in \mathbb{Z}\}$. The vacuum $\Phi_{2, vac} = 2 \theta_2$ was expected since it corresponds to the kink in (\ref{kink1c}). In addition, the vacua $\Phi_{2, vac}= 2\pi n_2$ are inherited from the original ATM model in (\ref{atm1}). For the boson mass one gets
\br\nonumber
m^2 &\equiv & - \frac{\pa^2 U^{(2)}[\Phi_2]}{\pa \Phi_2^2} |_{\Phi_2=0}\\ 
&=&  2 \kappa^2 (\sec{\frac{\theta_2}{2}})^2. \label{mass1c}
\er

Notice that the boson mass above has the same form as in the $n=1$ sector (\ref{mass1}). This was expected since these expressions are associated to the same elementary boson mass of the theory. On the other hand, the energy density associated to the scalar field in (\ref{phiestaticc}) $
{\cal \epsilon}(x) = \frac{1}{2} (\Phi_2')^2 +  U[\Phi]^{(2)}= 2U^{(2)}[\Phi]$,
provides the same soliton mass as in (\ref{solmass1c}) for the kink (\ref{kink1c}). However, the soliton parameters will be related below to the spinor parameters according to the eq. (\ref{cond2}), the boundary and normalization conditions.   

For later purpose let us consider the expression (\ref{solmass1c}) of the soliton mass and the boson mass (\ref{mass1c}) and write the relationship
\br
\label{msolmc}
M_{sol} &=& \{ 8 \sqrt{2}  [ 1- \theta_{2} \cot{\theta_{2}}] \cos{(\frac{\theta_{2}}{2})}\}\,\,m.
\er
This is an exact result and shows, as in (\ref{msolm}) for the $n=1$ sector, that the soliton mass is directly proportional to the boson mass. 

Next, we obtain the algebraic equation satisfied by the bound sate energy $E_2$. Substituting the relationships (\ref{tau01})-(\ref{tau4}) for $n=2$, given in terms of the tau functions (\ref{t01c})-(\ref{tauxi2c}), in the equations (\ref{5211})-(\ref{5241}) one gets  a homogeneous linear system of equations for the unknown independent parameters $\{ b_1, b_2,b_3,b_4\}$. Then imposing the condition of vanishing determinant for the $4 \times 4$ matrix formed by the coefficients in that linear system of equations one gets a quartic algebraic equation for the bound state energy $E_2$, corresponding to the both parities $\s=\pm 1$, respectively
 \br
  \label{quartic}
 r_4 E_2^4 &+& r_3 E_2^3 + r_2 E_2^2 + r_1 E_2 + r_0  - \s\, 4 \kappa\, M_1 (s_2 E_2^2 + s_1 E_2 + s_0) =0\er
where
\br
\nonumber
 r_4 &=& -8 M_1^2 (\sin{\theta_2})^4,\,\,\,r_3=64 \kappa M_1^2 \cos{\theta_2} (\sin{\theta_2})^3,\\
 r_2 &=& 128 \kappa^4 +
 2 M_1^2 \Big[  \kappa^2  - 2 M_1^2 -(12 \kappa^2 - \frac{7}{2} M_1^2) \cos{(2 \theta_2)} + (11 \kappa^2 - 2 M_1^2) \cos{(4 \theta_2)} + \nonumber \\
&& \frac{1}{2} M_1^2 \cos{(6 \theta_2)}\Big]\nonumber\\
 \nonumber
 r_1&=& 8 \kappa M_1^2 \Big[ 6 \kappa^2 + M_1^2 + (6 \kappa^2 - 2 M_1^2) \cos{(2 \theta_2)} +
   M_1^2 \cos{(4 \theta_2 )}\Big] \sin{(2 \theta_2 )}\\
   r_0 &=& 4 M_1^2 \Big[ 7 \kappa^4 - 5 \kappa^2 M_1^2 -
   M_1^4 + (9 \kappa^4 + 2 \kappa^2 M_1^2 + M_1^4) \cos{(2 \theta_2 )} +
   3 \kappa^2 M_1^2 \cos{(4 \theta_2 )}\Big] ( \sin{\theta_2} )^2 \nonumber\\
\nonumber
   s_2 &=& 2 \Big[ 8 \kappa^2 + M_1^2 - M_1^2 \cos{(2 \theta_2 )}\Big] \sin{(2 \theta_2 )},\,\,s_1=8 \kappa (\cos{\theta_2} )^2 \Big[ 2 \kappa^2 - M_1^2 + M_1^2 \cos{(2 \theta_2 )}\Big]\\\nonumber
   s_0 &=& M_1^2 \Big[\kappa^2 - M_1^2 + (3 \kappa^2 + M_1^2 ) \cos{(2 \theta_2 )}\Big] \sin{(2 \theta_2 )}.
 \er
In the Figs. 17 and  18 we provide some contour-plots for $E_2 \,\mbox{vs}\, \theta_2$ and in the Fig. 19  we plot $E_2 \,\mbox{vs}\, \rho_o$ for the special set of parameters.

The boundary condition for the spinor $\xi_1$ is given by
\br
\label{cond1c}
\xi_1(0)\equiv \xi_0 = \frac{1}{2} [\sec{(\frac{\theta_2}{2})}]^2 \( \rho_1 \cos{\a_1} + \rho_4 \s \,\sin{\a_4}\)
\er

The spinor normalization for a general set of parameters is provided in the appendix (\ref{normn2}). For a special choice of the parameters $\rho_a=\rho_o,\,\a_{a}=\a_o$ from (\ref{normn2}) one has
\br
\nonumber
\int_{-\infty}^{+\infty} \Big[ \sum_{a=1}^{4} \xi_{a}^2 \Big] dx &=& \frac{4 \rho_o^2 \s}{|\kappa| (\sin{\theta_2})^3}  \times\\
&& \Big\{ \theta_2  \s + \sin{(2
\a_o )} \sin{\theta_2}- \cos{\theta_2} [ \theta_2 \sin{(2
\a_o )}+ \s \sin{\theta_2} ]   \Big\}\label{cond31c} \\
&=&1, \label{cond32c}
\er
In the particular case $\a_o=0$ from (\ref{cond31c})-(\ref{cond32c}) one has
\br
\label{kappa2}
\frac{4 \rho_o^2}{ |\kappa|}  =  \frac{(\sin{\theta_2})^3}{ \theta_2 - \frac{1}{2} \sin{(2 \theta_2)} }
\er

\begin{figure}
\centering
\includegraphics[width=5cm,scale=1, angle=0,height=5cm]{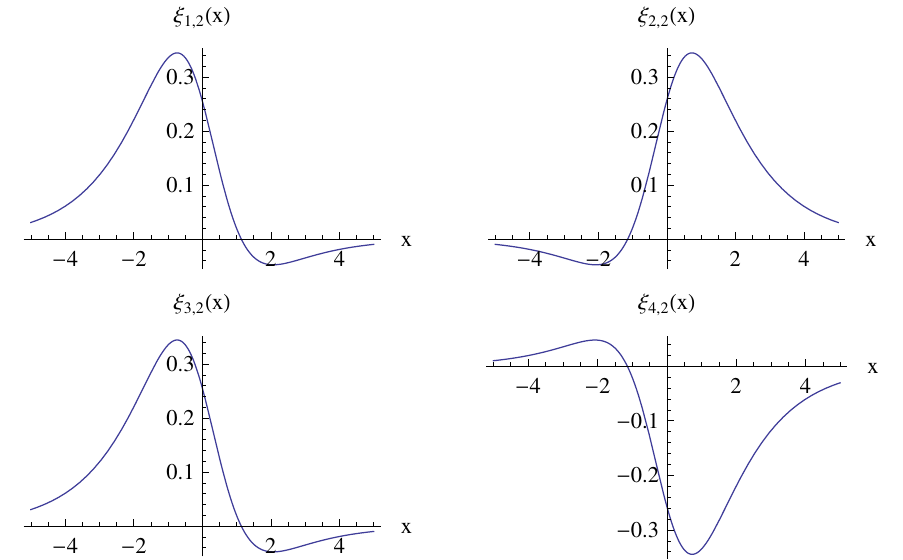}
\parbox{5in}{\caption{The bound state of the fermion $\xi_{a,2}$ as a function of $x$ for the {\bf positive parity} $\s=1$. It is plotted for $\rho_o = 0.5,\, \a_o =0,\, \theta_2 = 0.32,\, \kappa =0.69$ .  Notice the spinor components relationships $\xi_{1,2}(-x)= - \xi_{4,2}(x)$ and $\xi_{2,2}(-x)= + \xi_{3,2}(x)$}}
\end{figure}

\begin{figure}
\centering
\includegraphics[width=5cm,scale=1, angle=0,height=5cm]{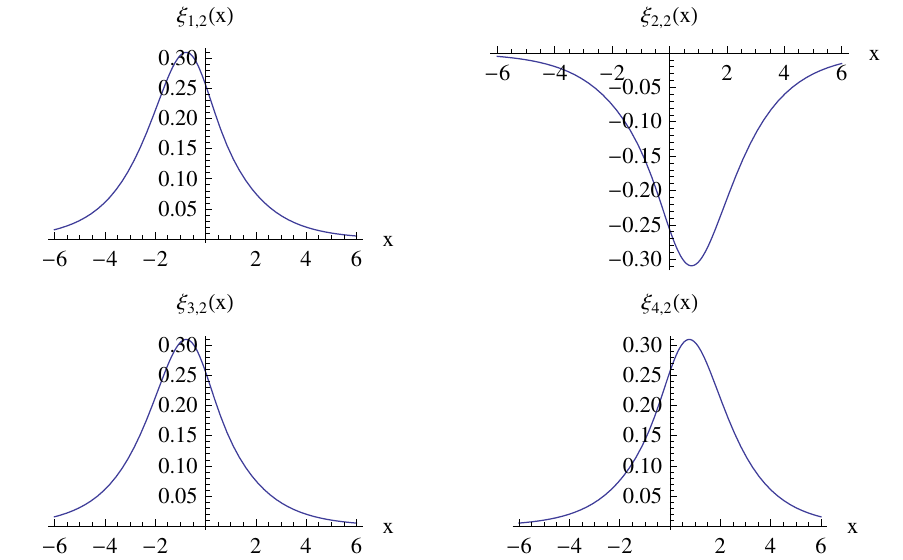}
\parbox{5in}{\caption{The bound state of the fermion $\xi_{a,2}$ as a function of $x$ for the {\bf negative parity} $\s=-1$. It is plotted for $\rho_o = 0.5,\, \a_o =0,\, \theta_2 = 0.32,\, \kappa =0.69$. Notice the spinor components relationships $\xi_{1,2}(-x)= + \xi_{4,2}(x)$ and $\xi_{2,2}(-x)= - \xi_{3,2}(x)$}}  
\end{figure}  

The  Figs. 12 and 13 show the solutions  $\xi_{a,2},\,\,a=1,...4$ defined in (\ref{tau1})-(\ref{tau4}) (for $n=2$) considering the tau functions (\ref{t01c})-(\ref{tauxi2c}) for $\s= \pm 1$, respectively. The parameters
in these Figs. satisfy the condition (\ref{cond2c}) and the normalization condition (\ref{cond31c})-(\ref{cond32c}). We must emphasize that the kink parameters $\kappa, \theta_2$ always
depend on the spinor parameters through the eqs. (\ref{cond2c}), the boundary condition (\ref{cond1c}) and the normalization condition (\ref{cond31c})-(\ref{cond32c}), so the shape of the kink always depends on the spinor parameters and vice versa.
    
In the  Fig. 14 it is plotted the function $ \frac{(\sin{\theta_2})^3}{\theta_2- \frac{1}{2} \sin{(2 \theta_2)}}>0$ which appears in (\ref{kappa2}) defining the regions of validity of $\theta_2$ consistent with the normalization condition (\ref{cond31c})-(\ref{cond32c}).

\begin{figure}
\centering
\includegraphics[width=5cm,scale=1, angle=0,height=5cm]{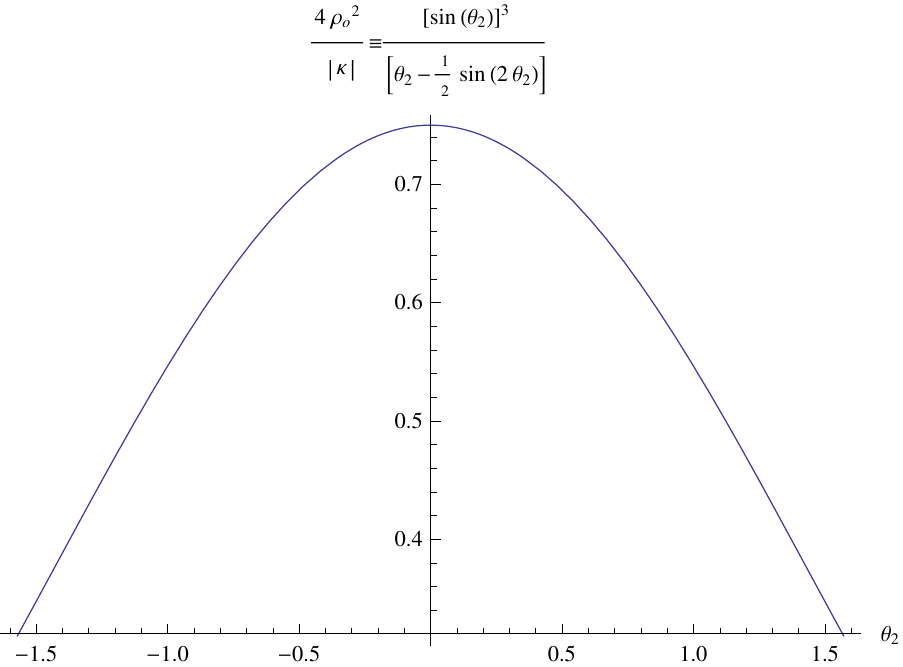}
\parbox{5in}{\caption{The graph of the function in (\ref{kappa2}) for $\theta_2 \in [-\pi/2\,,\,\pi/2]$ which must be equal to the ratio $\frac{4 \rho_o^2}{|\kappa|}$.}}
\end{figure}

\begin{figure}
\centering
\includegraphics[width=7cm,scale=2, angle=0,height=7cm]{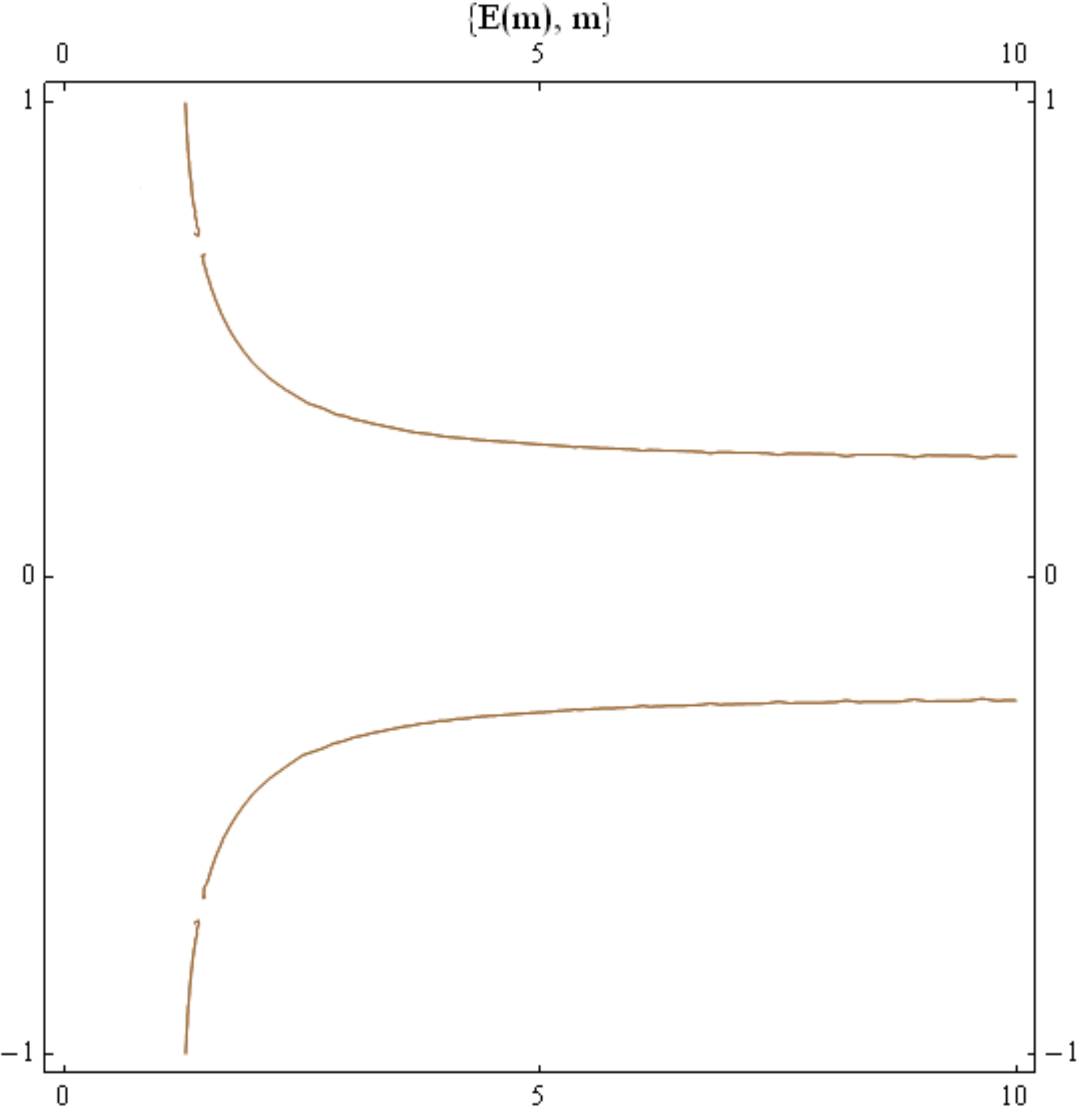}
\parbox{5in}{\caption{ The bound state energy of the fermion as a function of the boson mass $m$ for $M_1 = 1$ and $\theta_2 \rightarrow \pi/2$ for the both parities. The plot corresponds to either the positive or the negative parity. The curves never approach zero as $m$ increases, so these curves have the same behavior as the ones for the case $n=1$ (see Fig. 10).  }}
\end{figure}

\begin{figure}
\centering
\includegraphics[width=7cm,scale=2, angle=0,height=6cm]{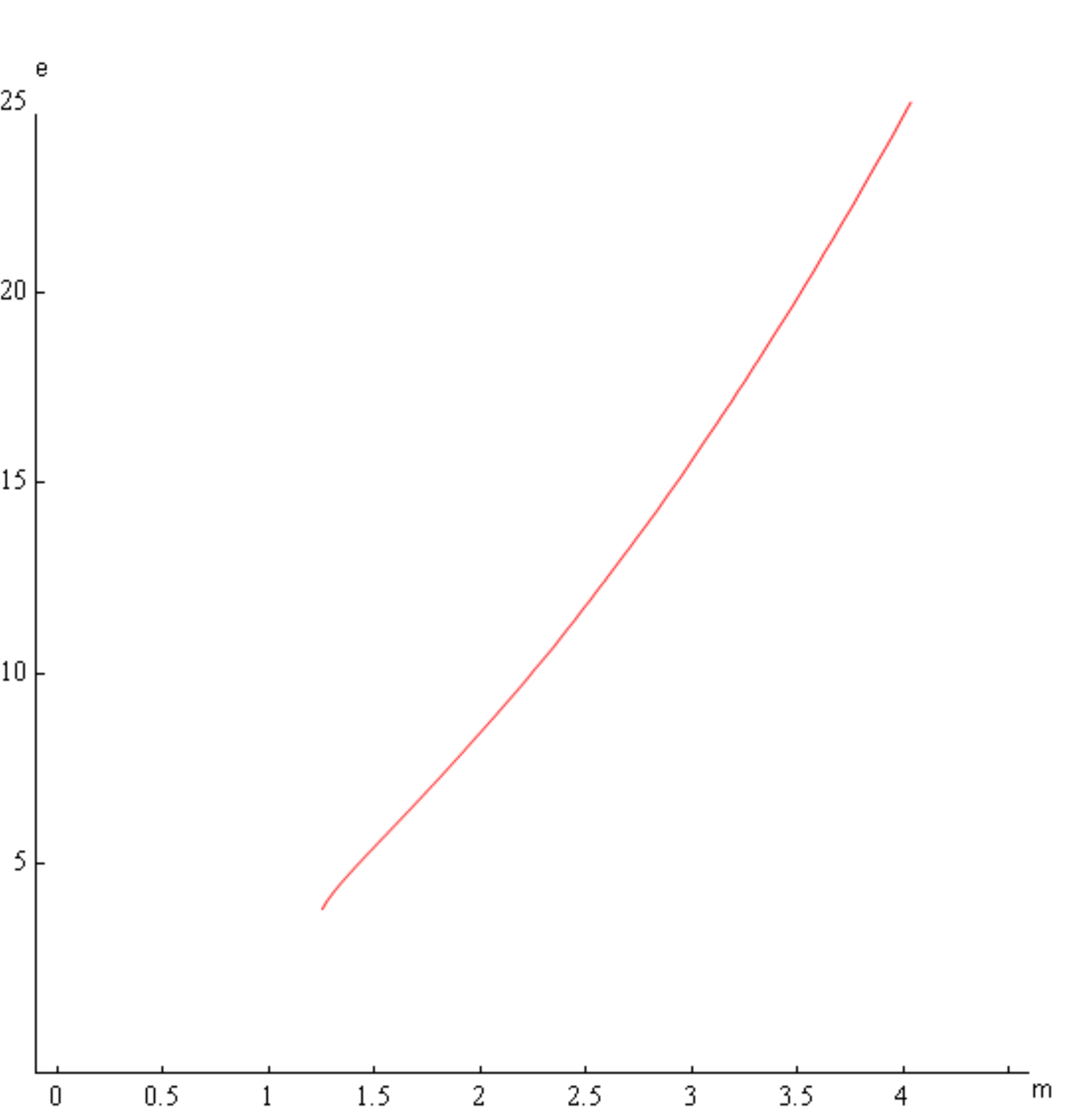}
\parbox{5in}{\caption{The total energy $e_2$ (soliton mass plus the fermionic bound sate energy) as a function of the boson mass $m$  for $M_1=1, \theta_2 \rightarrow \pi/2$. Notice that the line departs slightly from the linear behavior observed in the $n=1$ case (see Fig 9) in this limit. }}
\end{figure}

\section{Vacuum polarization}
\label{sec:pol}

The vacuum polarization of the fermion induced by
the presence of the soliton has been considered in \cite{goldstone} by introducing  a powerful method, called the adiabatic method. In this method, the topologically nontrivial configuration of the background scalar field
which is coupled to the fermions is assumed to evolve continuously and slowly from a
topologically trivial configuration. Later on in \cite{mackenzie} the adiabaticity requirement was lifted such that the energy spectrum of the fermions (in first quantization) coupled to the background field is computed during the adiabatic process. Defining the fermionic charge of the soliton as charge of the ground state, and  monitoring the energy spectrum, including the flow of any
level across zero, they have deduced that the vacuum polarization induced by the soliton receives two contributions. First is the adiabatic contribution, predicted in \cite{goldstone}, which is a linear
decrease with slope $- \phi_o/\pi, (\phi_o>0)$ whenever the interaction is in the form $\bar{\psi} e^{i \Phi \g_5} \psi,\,\,[\Phi(\pm \infty) = \pm \phi_{o} ]$. This is
due to the occurrence of the spectral deficiency in the Dirac sea as $\phi_o$ increases. The
second is the non-adiabatic contribution which occurs only when one or more fermionic energy levels
cross $E = 0$, since in that case the definition of ground state changes. In the case when one energy level crosses zero it must be considered a jump of $+1$ in the vacuum polarization since
this energy level is filled after crossing $E = 0$ in the vacuum state, by definition.
The formula for the vacuum polarization (VP) for the type of interaction described above becomes \cite{mackenzie,polychronakos,keil,gousheh}
\br\nonumber
\label{vacuumpol}
VP_{n} &=& - \frac{ \Big[ \Phi_{n}(+\infty)-\Phi_{n}(-\infty) \Big]}{2 \pi},
\,\,\,n=1,2,...\\
&=&- \frac{n \theta_n }{\pi},
\er
with possible jumps of $+1$ at configurations where level crossing occur. Notice that in the above eq. we have
considered the $VP_n$ for each topological sector $n=1,2,...,$ of the model considered in this paper.

In the Figs. 5 and 6 (case $n=1$) and 17 and 18 (case $n=2$) the energy spectrum of the fermions coupled to the background field is computed during the adiabatic process. For each plot it is possible to compute the vacuum polarization as the parameter $\theta_n$ increases using the formula (\ref{vacuumpol}) and considering the jump $+1$ every time a level crossing occurs.  

In the Figs. 7 and 8 (case $n=1$) and 19 ($n=2$) we have plotted the energy spectrum in terms of the spinor parameters $\rho_1\, (n=1),\, \rho_o\, (n=2)$, respectively,  for various fixed values $\Phi_n(\pm \infty)=n\theta_n\,(n=1,2)$ the solitons may achieve asymptotically. These plots show qualitatively the role of the local features of the solitons, such as the change of the slope at the origin by varying the parameters $\rho_o,\,\, \rho_1$, on the level crossing effects at $E_n=0$.   

\begin{figure}
\centering
\includegraphics[width=15cm,scale=4, angle=0,height=15cm]{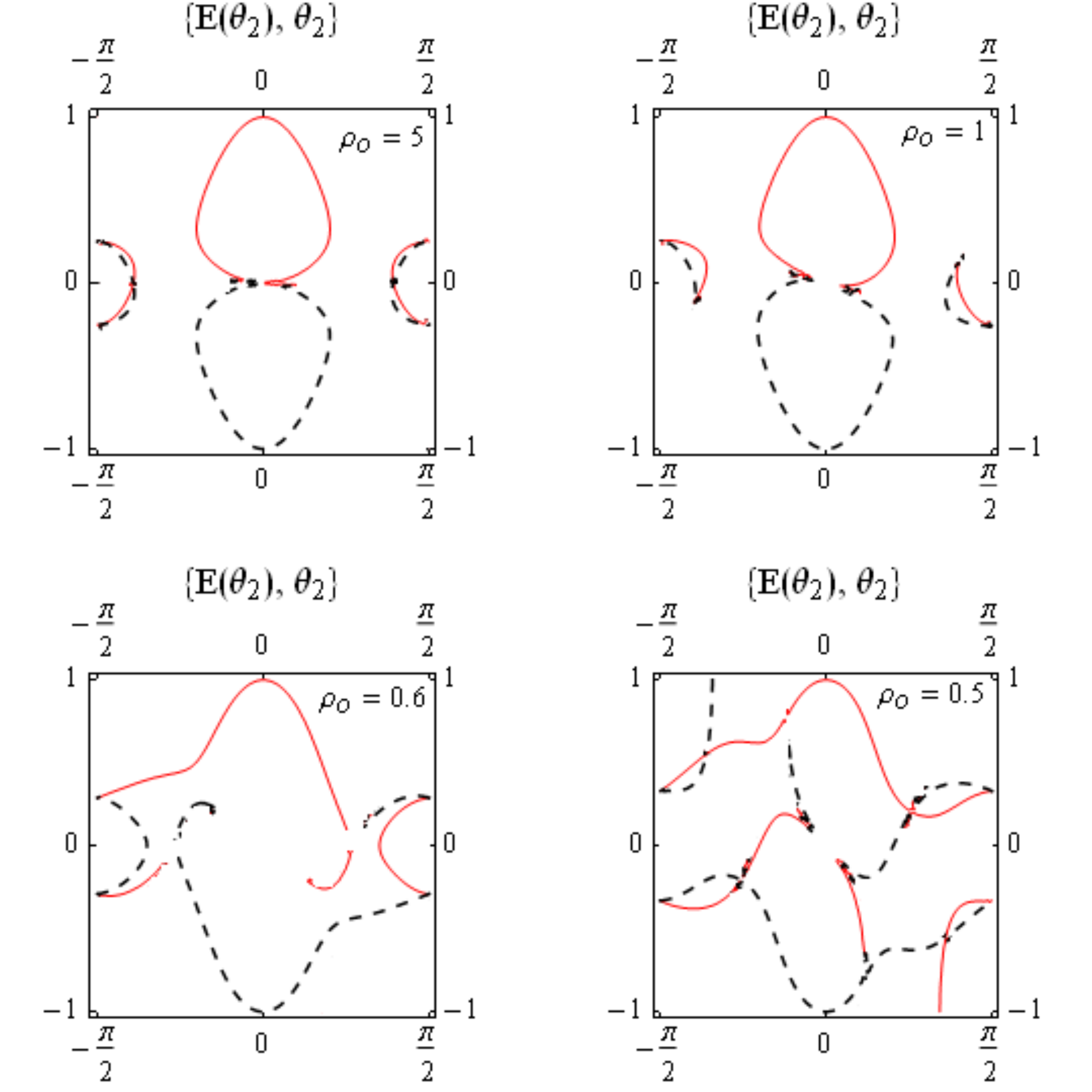}
\parbox{5in}{\caption{The fermionic bound state energy versus theta angle ($E\, vs\, \theta_2$) as a
contour plot for $M_1=1,\,\rho_{a}=\rho_0,\, \a_{a}=0,\, (a=1,2,..4)$ and the values $\rho_o = 5, 1, 0.6, 0.5$.  The continuous and the dashed lines correspond to $+$\,and $-$ parities, respectively. Notice that for $\rho_o \rightarrow +\infty$ there are some regions for $\theta_2$ with four and two bound states, respectively, whereas for some regions there is none.}}
\end{figure}

\begin{figure}
\centering
\includegraphics[width=15cm,scale=4, angle=0,height=15cm]{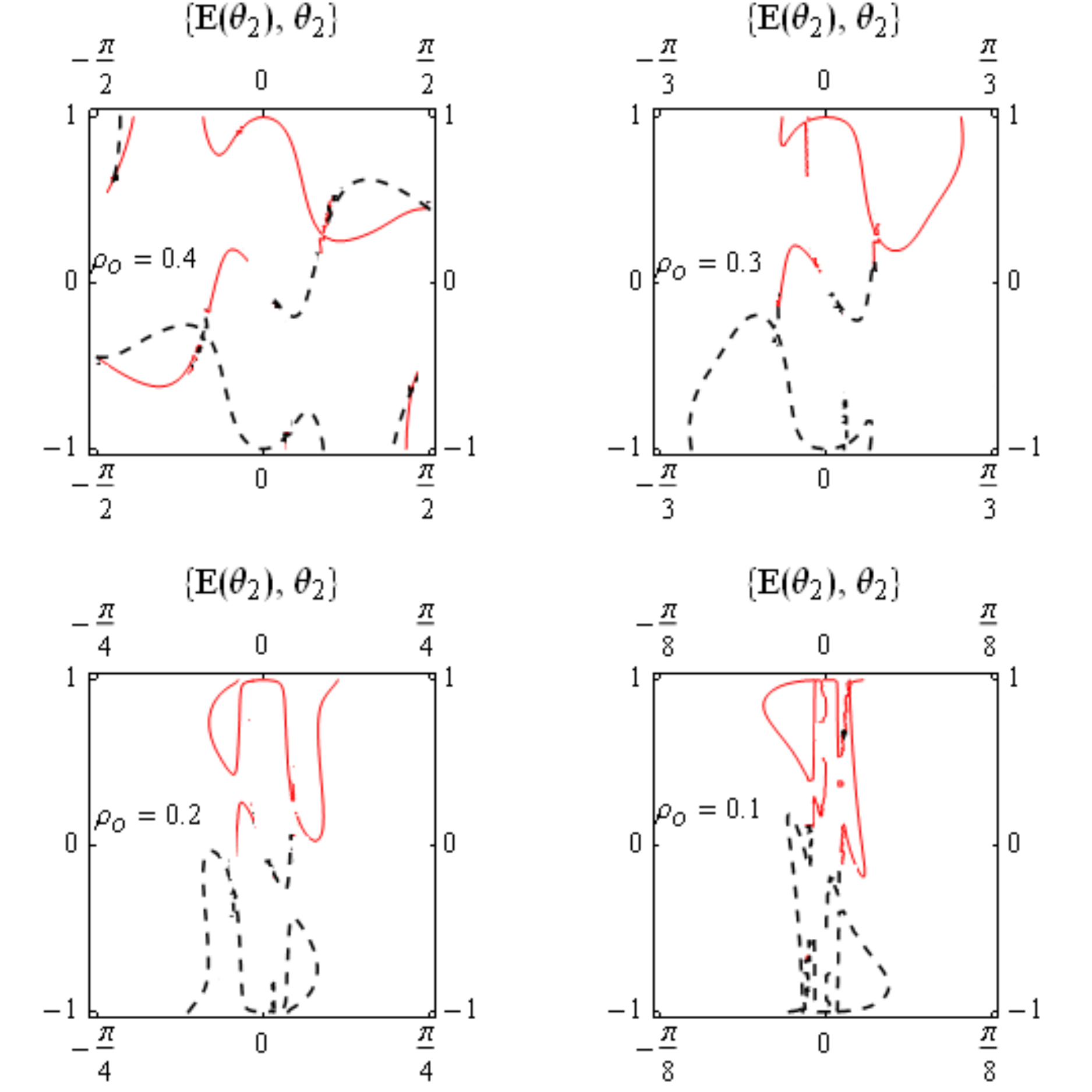}
\parbox{5in}{\caption{The fermionic bound state energy versus theta angle ($E\,\, vs\,\, \theta_2$) as a contour
plot for $M_1=1,\,\rho_{a}=\rho_0,\, \a_{a}=0,\, (a=1,2,..4)$, the values $\rho_o = 0.4, 0.3, 0.2, 0.1$ and the regions for $\theta_2 = \{[-\pi/2,\pi/2],[-\pi/3,\pi/3],[-\pi/4,\pi/4],[-\pi/8,\pi/8]\}$, respectively . The continuous and the dashed lines correspond to $+$\,and $-$ parities, respectively.  Notice that as $\rho_o$ decreases the regions for $\theta_2$ with bound states become smaller and close to the origin, whereas the regions with zero bound states becomes larger.}}
\end{figure}

\begin{figure}
\centering
\includegraphics[width=10cm,scale=1, angle=0,height=10cm]{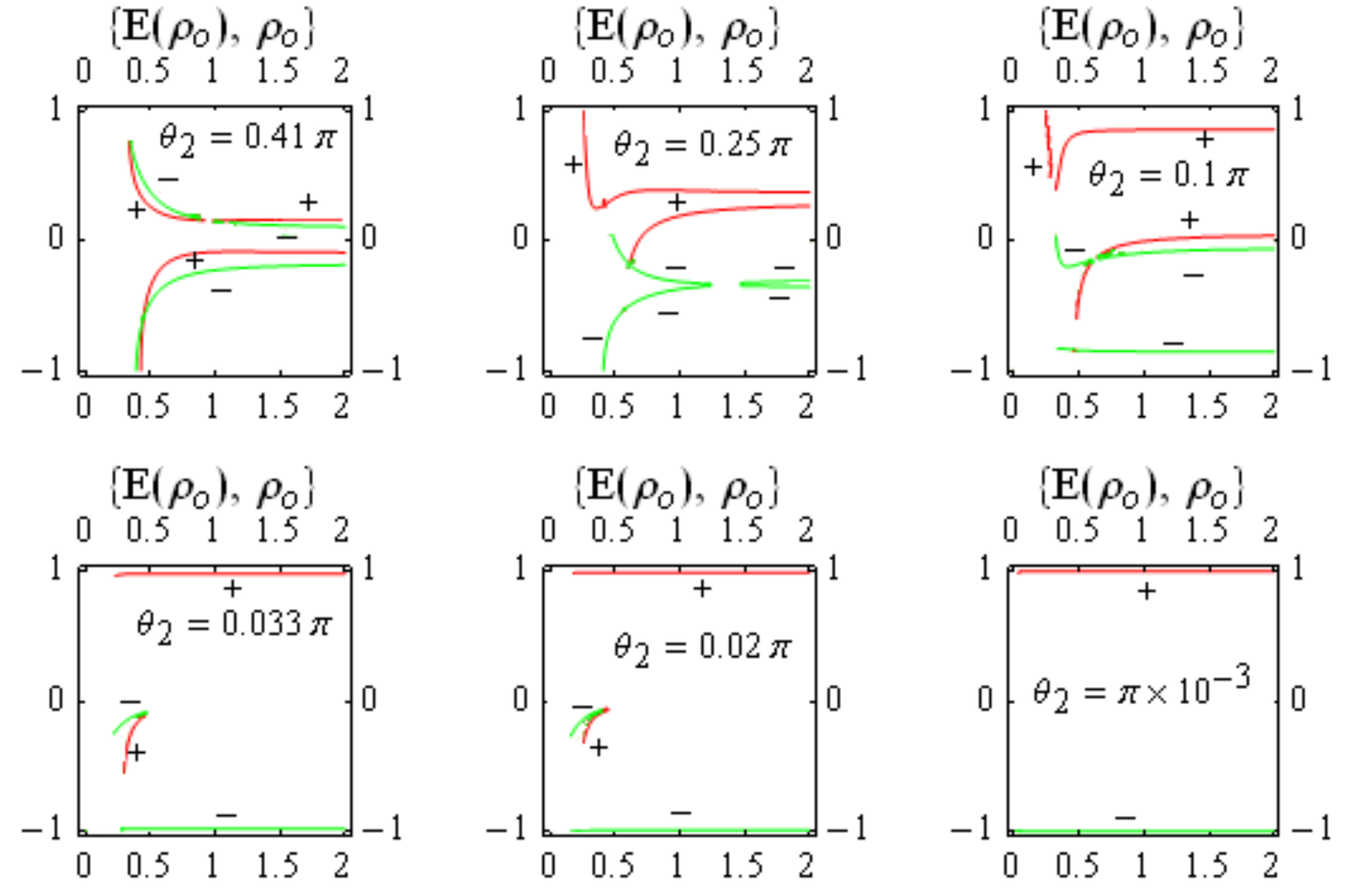}
\parbox{5in}{\caption{The fermionic bound state energies when $M_1=1,\,\rho_{a}=\rho_0,\, \a_{a}=0,\, (a=1,2,..4)$ and for the values $\theta_2 = 0.41 \pi, 0.25 \pi, 0.1 \pi, 0.033 \pi, 0.02 \pi, 10^{-3}\times \pi$, respectively.
 The  $+$\,and $-$ parity signs are indicated close to the corresponding lines, respectively. Notice that as $\theta_2$ decreases the positive (negative) parity bound states approaches the positive( negative) continuum and in general the number of bound states decreases.}}
\end{figure}

\section{Currents equivalence and charge quantization}
\label{sec:duality}

Finally, let us rewrite the currents equivalence (\ref{equivcurrents}) in terms of the scalar field $ \Phi $ and the spinors $\xi_{a}$; so, taking into account (\ref{zetas})-(\ref{zetas1}) one can write the zero component of the eq. (\ref{equivcurrents}) as
\br\label{equivcurrezetas}
 \frac{1}{2\pi} \frac{\pa}{\pa_x} (\Phi - \bar{\zeta} ) =\frac{2 e_{\psi}}{\pi k} \sum_{a=1}^{4} \xi_a^2.
\er

Taking into account $\bar{\zeta}=0$ and the spinor normalization condition to unity for any $n$ sector,
the cases $n=1,2$ are considered in (\ref{cond31})-(\ref{cond32}) and
(\ref{cond31c})-(\ref{cond32c}), respectively,  the integration of the relationship (\ref{equivcurrezetas}) provides
\br
\label{equivpar}
n \theta_n = \frac{2 e_{\psi}}{k},\,\,\,n=1,2,3...
\er
In this way we have established the relation between the angle $\theta_n$ and the parameters of the ATM model
$ k,  e_{\psi}$. Since $k$ is related to the WZNW model coupling constant $q \in \mathbb{Z}$ by $k = \frac{q}{2 \pi}$, then  from (\ref{equivpar}) one has the relationship
\br
\label{quanti1}
\frac{e_{\psi}}{\theta_n} = \frac{1}{4 \pi}\, n\, q;\,\,\,\,q \in \mathbb{Z}; \,\,n=1,2,...
\er

Moreover, the $VP_n\,$ expression can get another formulation in terms of the $e_{\psi}$ parameter and the WZNW coupling $q$, i.e.  $VP_{q} = - \frac{4 e_{\psi}}{q}$. Therefore, it is a remarkable fact that the topological and Noether currents equivalence present in the ATM model implies that the parameter $\theta_n$ defining the asymptotic values of the scalar $\Phi(\pm \infty)$ is related, in the spinor sector, to the parameter $e_{\psi}$ which defines the reality condition imposed on the two spinors of the ATM model (\ref{psitrealcond}).   

The relationship (\ref{quanti1}) has some remarkable properties. First, since the parameters $\theta_n$ and $e_{\psi}$ define the topological and spinor charges  through the eqs.  (\ref{topol11}), and  (\ref{psitrealcond}) and (\ref{noethersl2})-(\ref{psicharge}), respectively, together with (\ref{quanti1}) one can get 
\br
\label{equivch}
Q_{top}^{(n)} =  Q_{\psi}^{(q)},\,\,\,\,\,\,\,\,\,\,\,  Q_{top}^{(n)} \equiv \frac{n \theta_{n}}{\pi},\,\,\,Q_{\psi}^{(q)} \equiv \frac{4 e_{\psi}}{q}.\er 

It is clear that $Q_{top}^{(n)}$ may assume positive or negative values depending on the sign of $\theta_n$. Second, the relationship (\ref{quanti1}) shows that the ratio $e_{\psi}/ \theta_n $ must be quantized and that the $U(1)$ charge for fixed value of $q$ is quantized in units of the topological charge $Q_{top}^{(1)}$, i.e. $ Q_{\psi}^{(q)} =  \lb Q_{top}^{(1)} \rb\,\, n;\,\, Q_{top}^{(1)} = \frac{\theta_{o}}{\pi};\,\,\,(\theta_o \in [-\pi/2,\pi/2])$. Conversely, the topological charge $Q_{top}^{(n)}$ for fixed value of $n$  may assume discrete values, i.e. $ Q_{top}^{(n)} =  \lb Q_{\psi}^{(1)} \rb\,\, \frac{1}{q};\,\, Q_{\psi}^{(1)} \equiv 4 e_{\psi}$.

\section{Ratio (soliton mass)/(boson mass)}  
\label{sec:ratio}

Let us investigate the boson mass dependence of the kink energy in the cases $n=1, 2$.  Then from the relationships  (\ref{msolm}) and (\ref{msolmc}) one has for $\theta_n=\pi/2,\,n=1,2$ the ratios $M_{sol}/m$
\br
\label{ratioc}
\frac{M_{sol}}{m}  = \left\{ \begin{array}{ll} 2,  &\,\,\,\,\, n=1  \\
8, &\,\,\,\,\, n=2  \end{array}
\right.
\er
In the Fig. 22 this ratio has been plotted in terms of the angles $\theta_n\, (n=1,2)$. Notice
that for the soliton with $M_{sol}=1$ from (\ref{ratioc}) one has $m = 0.125$ in the $n=2$ topological sector, which is $\sim \,8\%$ smaller than the ratio between the experimental masses of the pion and the nucleon which is around $ \frac{\mbox{pion mass}}{\mbox{nucleon mass}} = 0.147$.
Therefore, if in the $n=2$ sector of the ATM model the pseudoscalar
field $\Phi(x)$ plays the role of the pions, solitonic excitations (with $M_{sol} = 1$ for $m=0.125$)
could be considered to be nucleons. In fact, our exact result improves the numerical result in \cite{Shahkarami1}
which provided $m_{numer.} = 0.116$, i.e. the numerical calculation result is about $7 \%$ less than the analytical result.

For $n=2$ case an implicit expression relating the bound state energy and the boson mass
can be obtained substituting  the relationship (\ref{mass1c}) for $\kappa$ into the eq. (\ref{quartic}). In Fig. 15
the plot $E$ vs $m$ is shown for $M_1=1$ in the limit $\theta_2 \rightarrow \pi/2$ for the both parities. In that figure  it can be seen that at $m=0.125$ there is no fermionic level crossing zero. Therefore, in this case the soliton of mass $M_{sol} = 1$ polarizes the vacuum and its ground state possesses fermion number one, the vacuum polarization being $(-1)$ for $\theta_2 = \pi/2$ according to the formula (\ref{vacuumpol}).

It could be interesting to consider the extension of the ATM model \cite{jhep1,jhep11, jhep2, bueno,jmp1} incorporating
more pions and nucleons to mimic the Skyrme model in order to partly account
for the difference with the experimental value of the above result regarding the ratio between the pion and nucleon.

\begin{figure}
\centering
\includegraphics[width=5cm,scale=1, angle=0,height=5cm]{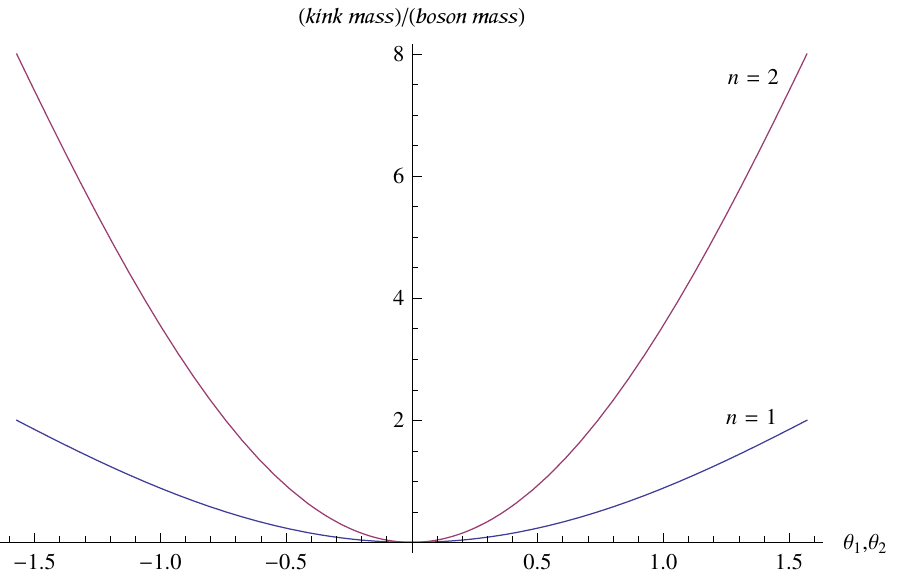}
\parbox{5in}{\caption{The ratio $\mbox{(kink mass)} / \mbox{(boson mass)}$ in the interval $\theta_1,\,\theta_2 \in [-\frac{\pi}{2}\,,\,\frac{\pi}{2}]$.}}
\end{figure}
  
\section{The topological sectors $n\geq 3$ }
  \label{sec:n3}

Following  the previous constructions one can extend to the case $ n \geq 3 $. The tau functions in (\ref{tau01})-(\ref{tau4}) can be written as
\br
\label{tau0n}
\tau_0^{(n)} &=& 1+ e^{-i \theta_n} e^{2 \kappa x},\,\,\,\, \tau_1^{(n)}=[\tau_0^{(n)}]^\star\\
\label{tau1n}\tau_{\xi,\,n}^{a} &=& b_a e^{\kappa x} + b_{aa} e^{3 \kappa x}+...+ b_{\underbrace{aa...a}_{``n"times}} e^{[(2 n -1) \kappa x]} ,\,\,\,\,a=1,2,..4\\
\widetilde{\tau}_{\xi,\,n}^{a} &=& \bar{b}_a e^{\kappa x} + \bar{b}_{aa} e^{3 \kappa x}+...+ \bar{b}_{\underbrace{aa...a}_{``n" times}} e^{[(2 n -1) \kappa x]}.\label{tau2n}
\er

Let us discuss the appearance of the algebraic equation for the eigenvalue $E$ in the case $n\geq 3$, the analog of the quadratic (\ref{quadratic1}) and quartic (\ref{quartic}) equations  in the cases $n=1,2$, respectively.
Since the eqs. (\ref{5231})-(\ref{5241}) are related to the eqs. (\ref{5211})-(\ref{5221})  by parity transformation it is enough to consider the last couple of eqs. in order to get the mentioned algebraic equation for the eigenvalues  $E$. Substituting the relationships (\ref{tau01})-(\ref{tau4}) given in terms of the tau functions (\ref{tau0n})-(\ref{tau2n}) into the equations (\ref{5211})-(\ref{5221}) will provide a $2n-$order algebraic equation for the bound state energy $E$. In fact, notice that the parameter $E$ appears as a coefficient of the fields $\xi_1$ and $\xi_2$  in (\ref{5211})-(\ref{5221}), and this system of eqs. will provide a homogeneous linear system of eqs. for the unknown parameters $b_a,\,b_{aa},...b_{\underbrace{aaa...a}_{``n" times}}, \,\bar{b}_a,\,\bar{b}_{aa},...\bar{b}_{\underbrace{aaa...a}_{``n" times}},\,(a=1,2)$. Moreover, the parameters of type  $\bar{b}_{aa...a}$ depend on the other set of parameters  $b_{aa...a}$ through the parity relationships (\ref{parity}) [in the cases $n=1,2$ one has the eqs. (\ref{barb1}) and (\ref{relations11})-(\ref{relations12}), respectively ]. So, in view of the expressions  of the eqs. (\ref{5211})-(\ref{5221})  in terms of the relevant tau functions, the eigenvalue parameter $E$ will appear only in the coefficients of the independent parameters $\Big[b_1,\,b_{11},...\, b_{\underbrace{11...1}_{``n" times}} $ $b_2,\, b_{22},\,... b_{\underbrace{22...2}_{``n" times}}\Big]$.  Then, in order to get a non-trivial solution one must impose a condition $det[A_{pq}] = 0$, where the matrix components $A_{pq}\,\,(p,q=1,...,2n)$ are the coefficients of the  $2n-$order homogeneous  linear system of eqs. for the above independent parameters. The outcome will be an algebraic equation of order $2n$ for $E$ with coefficients depending on the parameter constants $\(\kappa,\, M_1,\, \s,\, \theta_n,\)$.

Finally, if one considers the eqs. (\ref{quadratic1}) and (\ref{quartic}), respectively, as plane curves of fourth and sixth degrees in the variables $\{E, \kappa\}$ we can imagine that the possible combinations of coefficients (depending on the parameters $M_1, \theta_n, \s$) in the relevant algebraic equations will give rise to various families of plane curves; in this way the problem of the spectral flow of the fermion-soliton model due to the variations of the local features of the soliton, considered here for each topological sector $``n"$, will be connected to the study of the plane curves of $2(n+1)$ degree in the realm of algebraic geometry (see e.g. \cite{wolfram}). Notice that $\kappa$ is related to the soliton and fermion bound state parameters through the normalization conditions like (\ref{cond31})-(\ref{cond32}) and (\ref{cond31c})-(\ref{cond32c}), and this parameter determines the local properties of the soliton such as the slope at the origin for each value of $\theta_n$, as in (\ref{slope1}) and (\ref{slope2}).

\section{Discussion}
\label{sec:discussion}
We have considered special type of fields in the affine Toda model coupled to matter providing the analytical solutions as bound states of the fermion and solitons of the scalar field. The solutions of a static version of the system are obtained by using the tau function approach which has been previously proposed in the relativistic version of the theory. The self-consistent solutions appear as an infinite number of topological sectors labeled by $n \in \mathbb{Z}_{+}$, such that in each sector the scalar field would evolve continuously from a trivial configuration to the one with half integer topological charge $\pm \frac{n}{2}$. The spinor bound states are found analytically for each topological configuration of the background scalar field. In these developments the soliton shape depends crucially on the spinor bound state parameters, i.e. it depends on the fermionic state to
which it is coupled.  It is worth mentioning that we have considered the back-reaction of the spinor on the soliton exactly by using our analytical methods, this may be contrasted to the previous results in the literature in which the spinor back-reaction has been neglected or either considered perturbatively.  

The bound state energy in each topological sector satisfies an algebraic equation of degree $2n$ with coefficients depending  on the parameters $\kappa, M_1, \s, \theta_n$,  so the study of the  energy spectrum finds a connection to the realm of algebraic geometry. We provided explicit computations for the topological sectors $n=1,2$. Then, by monitoring the energy spectrum, including the energy flow of any level across $E_n=0$, we have provided the formula in order to compute  the vacuum polarization induced by the soliton. In this connection we have displayed some figures in the both topological sectors and elaborated on the adiabatic and non-adiabatic contributions to the vacuum polarization. 

It is shown that the equivalence between the Noether and topological currents present in the ATM model and the fact that the coupling constant is related to the integer coupling of the Wess-Zumino-Novikov-Witten (WZNW) model imply the spinor charge quantization.  

Considering the spinor fields as functionals of the scalar field we have found their contribution to the potential associated to the scalar field, thus computing a soliton mass and the relevant elementary boson mass. In the $n=2$ topological sector we have determined that the ratio  $m/M_{sol}$ approaches $(\mbox{boson mass}) /(\mbox{soliton mass}) \approx 0.125$, an error of about $8\%$ with respect to the 4D experimental result which is around $(\mbox{pion mass})/(\mbox{nucleon mass}) \approx 0.147$, in this way  reproducing the conjecture of Skyrme. So, we have extended the results of ref. \cite{Shahkarami1} concerning several properties of the bound states and improved the result regarding the Skyrme's conjecture. The construction of the bound states corresponding to the topological sectors $n \geq 3$ was briefly outlined. 

Finally, in \cite{matter} the authors discussed a linear dependence on the fermion charge $Q_{\psi}$ of  the both soliton and spinor particle masses,  $M^{'}_{soliton} \sim Q_{\psi} $, $\,M^{'}_{\psi} \sim Q_{\psi}$; however, the solitons discussed there were of the type (\ref{sol1}) associated to the  vacua $\vp_{vac.} = \pi n_1,\,\, n_1 \in \mathbb{Z},\,\psi_{vac.}=0$, and the spinor particle mass was defined by expanding the interaction term in (\ref{atm1}) around that soliton vacua. It could be interesting to discuss such relationships in the framework of our analytical solutions,  in a future investigation. We may argue that those type of relationships are concerned with the correspondence (spinor) particle/soliton in the ATM theory, indicating some sort of duality similar to the electromagnetic duality of some four-dimensional gauge theories possessing the
Bogomolny (monopole) limit \cite{bogomolny}, where the masses of particles and
monopoles (dyons) are given by $mass  \sim  \sqrt{Q_{elect.}^2 + Q_{mag.}^2 }$ .

\vspace{1cm}

\noindent {\bf Acknowledgements}

The author would like to thank PRONEX-CNPq-FAPEMAT for partial financial support and J. M. Jaramillo for discussions.

 \appendix
\section{Spinor normalization and the functionals  $U_{1,\,2}^{(n)}[\Phi_n]$ for $n=2$.}
\label{app:norman2}

We present some results related to the topological sector $n=2$. The normalization of the spinor function in terms of the full set of parameters becomes
\br
\nonumber
\int_{-\infty}^{+\infty} \Big[ \sum_{a=1}^{4} \xi_{a}^2 \Big] dx &=& \frac{1}{12 |\kappa| \s (\sin{\theta_2})^3} \times 
\\
\nonumber
&& \Big\{
12 \theta_2 ( \rho_1^2 + \rho_2^2 + \rho_3^2 + \rho_4^2) \sigma -24 \theta_2 \cos{\theta_2} [ \rho_2 \rho_3 \sin{(\alpha_2 +
\alpha_3)} +\\
\nonumber
&& \rho_1 \rho_4 \sin{(\alpha_1 + \alpha_4)}] + 24 \sin{\theta_2} [ \rho_2 \rho_3 \sin{(\alpha_2 + \alpha_3)} +
    \rho_1 \rho_4 \sin{(\alpha_1 + \alpha_4)} ] + \\
&& \nonumber
8 (\sin{\theta_2})^3 [ \rho_2 \rho_3 \sin{(\alpha_2 - \alpha_3)} - rho_1 \rho_4 \sin{(\alpha_1 - \alpha_4)} ] +  \\ \nonumber && 8 \sigma (\sin{\theta_2})^4 [ \rho_1^2 \sin{(2 \alpha_1)} -  \rho_2^2 \sin{(
2 \alpha_2)} + \rho_3^2 \sin{(2 \alpha_3)} - \rho_4^2 \sin{(2 \alpha_4)}]- \\ \nonumber 
&& 2 \sigma \sin{(2 \theta_2)} [3 (\rho_1^2 + \rho_2^2 + \rho_3^2 + \rho_4^2) - \\
\nonumber
&& \rho_1^2 \cos{(2 \alpha_1)} + \rho_2^2 \cos{(
   2 \alpha_2)} - \rho_3^2 \cos{(2 \alpha_3)} + \rho_4^2 \cos{(
   2 \alpha_4)} ]+ \\
&& \sigma \sin{(4 \theta_2)} [ -\rho_1^2 \cos{(2 \alpha_1)} + \rho_2^2 \cos{(
     2 \alpha_2)} - \rho_3^2 \cos{(2 \alpha_3)} + \rho_4^2 \cos{(
     2 \alpha_4)}]
\Big\}.\nonumber \\ \label{normn2}
\er

Next, we provide the general form of the functionals $U_{1}^{(2)}[\Phi_2]$ and $U_{2}^{(2)}[\Phi_2]$ in terms of the full set of  parameters. Considering $R[\Phi_2]= \frac{\sin{[\frac{1}{4}(2 \theta_2 + \Phi_2)}]}{\sin{[\frac{1}{4}(2 \theta_2 - \Phi_2)}]}$ one has
\br\nonumber
U_{1}^{(2)}[\Phi_2] &=& \frac{1}{2} R[\Phi_2 ] \times \\ && 
\nonumber \Big\{ \frac{e^{(2 i \theta_2)} M_2 ( e^{i \theta_2} + R[\Phi_2])^2 [ \bar{b}_1 -
   \bar{b}_2 + (\bar{b}_{11} - \bar{b}_{22}) R[\Phi_2] ] [\bar{b}_3 -
   \bar{b}_4 + (\bar{b}_{33} - \bar{b}_{44}) R[\Phi_2] ]}{(1 + e^{i\theta_2} R[\Phi_2] )^6} + \\
\nonumber
&& \frac{e^{(2 i \theta_2)} M_2 (1+ e^{i \theta_2} R[\Phi_2])^2 [ b_1 -
   b_2 + (b_{11} - b_{22}) R[\Phi_2] ] [b_3 -
   b_4 + (b_{33} -b_{44}) R[\Phi_2] ]}{( e^{i\theta_2}+ R[\Phi_2] )^6}+\\
\nonumber
&& \frac{e^{(2 i \theta_2)} M_2  [ \bar{b}_1 -
   \bar{b}_2 + (\bar{b}_{11} - \bar{b}_{22}) R[\Phi_2] ] [b_3 +
   b_4 + (b_{33} + b_{44}) R[\Phi_2] ]}{(1+ e^{i\theta_2}R[\Phi_2] )^4}  +\\
\nonumber
&& \frac{e^{(2 i \theta_2)} M_2  [ b_1 +
   b_2 + (b_{11} + b_{22}) R[\Phi_2] ] [\bar{b}_3 -
   \bar{b}_4 + (\bar{b}_{33} - \bar{b}_{44}) R[\Phi_2] ]}{(1+ e^{i\theta_2}R[\Phi_2] )^4}+\\
\nonumber
&&\frac{e^{(-2 i \theta_2)} M_2  [ \bar{b}_1 +
   \bar{b}_2 + (\bar{b}_{11} + \bar{b}_{22}) R[\Phi_2] ] [b_3 -
   b_4 + (b_{33} - b_{44}) R[\Phi_2] ]}{(1+ e^{(-i\theta_2)}R[\Phi_2] )^4}  +
\\
\nonumber
&&
\frac{e^{(-2 i \theta_2)} M_2  [ b_1 -
   b_2 + (b_{11} - b_{22}) R[\Phi_2] ] [\bar{b}_3 +
   \bar{b}_4 + (\bar{b}_{33} + \bar{b}_{44}) R[\Phi_2] ]}{(1+ e^{(-i\theta_2)}R[\Phi_2] )^4}+\\
   \nonumber
   &&
   \frac{M_2 [b_1 + b_2 + (b_{11} + b_{22}) R[\Phi_2]] [b_3 +
     b_4 + (b_{33} + b_{44}) R[\Phi_2]]}{|1 + e^{(i \theta_2)} R[\Phi_2] |^4}+\\
\nonumber
&& \frac{M_2 [\bar{b}_1 + \bar{b}_2 + ( \bar{b}_{11} + \bar{b}_{22}) R[\Phi_2] ] [ \bar{b}_3 +
    \bar{b}_4 + (\bar{b}_{33} + \bar{b}_{44} ) R[\Phi_2]]}{|1 + e^{(i \theta_2)} R[\Phi_2] |^4}+
\\
&& \frac{16 i e^{(-i \theta_2)} [-1 + e^{(2 i \theta_2)}] \kappa^2 (\csc{[
  \frac{1}{4} (2 \theta_2 - \Phi_2)]})^2 \sin{\theta_2} \sin{\frac{
\Phi_2}{2} )}}{|1 + e^{(i \theta_2)} R[\Phi_2] |^4}
 \Big\}
\label{u1n22}
\er
and 
\br
U_{2}^{(2)}[\Phi_2 ] & = & \frac{1}{2} \frac{e^{(-4 i \theta_2 )} M_2 R[\Phi_2 ]}{|1 + e^{(i \theta_2 )} R[\Phi_2 ] |^6 }
 \times  \nonumber \\
&& \nonumber
\{ - ( e^{(i \theta_2 )} + R[\Phi_2] )^8 [ ( b_{11} - b_{22} + (b_1 - b_2) R[\Phi_2 ] ) ( b_{33} -
   b_{44} + (b_3 - b_4) R[\Phi_2 ] ) ] + \\
&& \nonumber [ (b_1 - b_2 + ( b_{11} - b_{22}) R[\Phi_2 ] ) ( b_3 - b_4 + (b_{33} - b_{44} ) R[\Phi_2 ]) ] (1 +
   e^{(i \theta_2 )} R[\Phi_2 ])^8-
\\
&& \nonumber
[(b_1 + b_2) (b_3 + b_4) - (b_{11} + b_{22}) (b_{33} + b_{44} )] e^{(4 i \theta_2 )} |1 + e^{(i \theta_2 )} R[\Phi_2]|^4 (-1 + R[\Phi_2]^2 )
-\\
&&
\nonumber
i \s e^{(2 i \theta_2)} |1 + e^{(-i \theta_2)} R[\Phi_2]|^2  (1+e^{(i \theta_2)} R[\Phi_2])^4 [ (b_{11} + b_{22} +
(b_1 + b_2) R[\Phi_2 ] ) \times \\
&&
\nonumber (b_1 - b_2 + (b_{11} - b_{22}) R[\Phi_2 ] ) +
 (b_{33} + b_{44} + (b_3 + b_4) R[\Phi_2]) (b_3 - b_4 + (b_{33} - b_{44}) \times \\
&& \nonumber  R[\Phi_2]) ]+
 i \s e^{(6 i \theta_2)} |1 + e^{(i \theta_2)} R[\Phi_2]|^2  (1+e^{(-i \theta_2)} R[\Phi_2])^4 [(b_{11} - b_{22} + (b_1 - b_2) \times \\
 && R[\Phi_2 ] ) (b_1 + b_2 + (b_{11} + b_{22}) R[\Phi_2 ] ) +
 (b_{33} - b_{44} + (b_3 - b_4) R[\Phi_2]) (b_3 + b_4 +  \nonumber \\
 && (b_{33} + b_{44}) R[\Phi_2]) ]
\} \label{u2n22}
\er

\end{document}